\documentclass[%
 reprint,
 amsmath,amssymb,
 aps,
]{revtex4-2}

\usepackage{graphicx}
\usepackage{dcolumn}
\usepackage{bm}
\usepackage{algorithm}
\usepackage{algpseudocode}

\newtheorem{theorem}{Theorem}

\begin{document}

\preprint{APS/123-QED}

\title{Resource-rational reinforcement learning and sensorimotor causal states, \\ and resource-rational maximiners}

\author{Sarah Marzen}
\affiliation{%
 Department of Natural Sciences\\
 Pitzer and Scripps College
}%




\date{\today}

\begin{abstract}
We propose a new computational-level objective function for theoretical biology and theoretical neuroscience that combines: reinforcement learning, the study of learning with feedback via rewards; rate-distortion theory, a branch of information theory that deals with compressing signals to retain relevant information; and computational mechanics, the study of minimal sufficient statistics of prediction also known as causal states. We highlight why this proposal is likely only an approximation, but is likely to be an interesting one, and propose a new algorithm for evaluating it to obtain the newly-coined ``reward-rate manifold''. The performance of real and artificial agents in partially observable environments can be newly benchmarked using these reward-rate manifolds. To that end, we calculate an example reward-rate manifold utilizing new equations reminiscent of the Blahut-Arimoto algorithm and propose a general algorithm for computing reward-rate manifolds. Finally, we describe experiments that can probe whether or not biological organisms are resource-rational reinforcement learners, using as an example maximin strategies, as bacteria have been shown to be approximate maximiners-- doing their best in the worst-case environment, regardless of what is actually happening. This proposal explains why ``good-enough'' for organisms might actually be near-optimal, if viewed correctly.
\end{abstract}

\maketitle

\section{Introduction}

According to Marr, understanding biological organisms entails uncovering three levels: the computational, the algorithmic, and the mechanistic \cite{marr1982vision, levenstein2023role}. At the computational level, we ask what organisms are trying to do. What objective function might they be using? At the algorithmic level, we ask what algorithm they are using to accomplish that objective. And at the mechanistic level, we ask how they are implementing that algorithm in their wetware. None of these levels have been completely understood in theoretical neuroscience or theoretical biology, despite major advances such as the Hodgkin-Huxley model that describes how neurons behave using electrical engineering ideas and the efficient coding hypothesis \cite{barlow1961possible} that describes how the brain has adapted to naturalistic stimuli.

In this manuscript, we claim that resource-rational decision making is a plausible first attempt at the computational level \cite{icard2023resource}, giving an optimization approach to biology. This research program goes by the name of computational rationality \cite{gershman2015computational}, rational inattention \cite{sims2003implications,sims2006rational}, and many other names. The basic idea behind it is that organisms endeavour to solve tasks as well as possible, but are limited in their ability to solve tasks by various resources. These resources can be time limitations, memory limitations, material limitations, or other limitations.

In this outlook, organisms that have been merely described as ``good enough'' at a task in the past are instead rationally inattending to certain bits of information, such that they are doing the best that they can with limited resources and essentially satisficing near-optimally. In other words, this research program asserts that \emph{given their resources}, a wide variety of organisms are doing the best that they can at gathering reward.

There is much debate over how to implement resource-rational decision making quantitatively, but information-theoretic codings of resources \cite{berger} and reinforcement learning-based measures of the quality of decision making \cite{sutton2018reinforcement} might be the key to understanding the full sensorimotor loop. Already, reinforcement learning has been famously used to describe dopaminergic signals \cite{schultz1997neural}, although there is much recent debate over whether or not that mechanistic level description is appropriate \cite{jeong2022mesolimbic}. On the other hand, using information-theoretic quantities as perceptual costs has allowed researchers to explain a number of empirical findings in a wide variety of areas in the last two decades, including various aspects of macroeconomic behavior \cite{sims2003implications,sims2006rational}, Shepard's universal law of generalization \cite{sims2018efficient}, the fuzziness of color naming systems \cite{zaslavsky2018efficient}, sub-optimal prediction in sequence learning \cite{humanbehavior}, and a number of empirical findings on neural coding and working memory \cite{jakob2023rate}. And, while not done on humans, recent work has shown that salamander retinal ganglion cells \cite{palmer2015predictive} and cultured cortical neurons from rats \cite{pnasnexus} both predict stimuli efficiently in an information-theoretic sense but do not always predict well in an absolute sense. Information-theoretic costs can be justified both using material constraints \cite{berger} and nonequilibrium thermodynamics \cite{hasenstaub2010metabolic, mehta2012energetic}.

There have been attempts to combine information-theoretic resource constraints and reinforcement learning objectives in Refs. \cite{still2009information,lai2023human,malloy2021capacity}, but in this manuscript, we will argue that these attempts require combination to achieve the correct objective. We will give a new Blahut-Arimoto-like algorithm for calculating what we call the ``reward-rate manifold'', which describes how well an organism (real or artificial) can attain reward under the information-theoretic resource constraints. In order to provide an algorithm, we will prove that the sensorimotor causal states of Ref. \cite{still2009information} can replace semi-infinite histories of observations and actions, essentially making it possible to calculate an infinite object with finite resources. As a corollary, it becomes clear that in the limit of resource constraints being inessential to functioning, sensorimotor causal states are stored by the organism. We then use a Gaussian Infomation Bottleneck-like \cite{chechik2003information} take on the equations underlying this algorithm to compute the reward-rate manifold for a simple example, showing that indeed, this computational-level objective function can be tested empirically with enough compute resources. We propose to use these kinds of analyses for more complicated systems, both computationally and experimentally, therefore testing if organisms are resource-rational decision makers, and if so, what kind.

We begin by describing the new proposed objective function and continue by providing an algorithm to efficiently calculate the newly-described reward-rate manifold and an example thereafter. We move to discussion of what else organisms might be doing-- maximizing their reward in the worst-case scenario. We conclude by describing what might be done in theoretical biology and even in machine learning (for evolved and engineered organisms) with this contribution.

Key to this research program is the ambitious idea that actually, organism brain and behavior all largely roughly obey the same objective function: Changes between organisms come from changes in their environments and their allowed level of resources. The exception to this comes lower-level organisms, like bacteria, that fail to have a theory of mind that would allow them to exploit the environment more intelligently and dynamically.

\section{A New Computational-Level Objective for Theoretical Biology}

We start by discussing proposals for a computational-level objective for theoretical biology in Sec. \ref{sec:1} and move to introducing my own in Sec. \ref{sec:2}. The environment under consideration is known in reinforcement learning \cite{sutton2018reinforcement} as a Partially Observable Markov Decision Process (POMDP), in which there is an underlying Markov state $w$ describing the environment, actions $a$ that describe what the agent can do, noisy and partial observations $o$ of the underlying world state $w$ that describe what the agent sees, a discount factor $\gamma$ that describes how agents treat future rewards, and a reward function $r(w,a)$ that describes how much ``reward'' an agent receives when the world is in state $w$ and the agent takes action $a$. These rewards can take the form of food, shelter, sleep, and so on, and are left unspecified for the purpose of this paper. In an experiment, one might imagine giving rats sugar or humans money. Mathematically, we specify $p(w_{t+1}|w_t,a_t)$ to be the way in which the organism's actions affect how the world evolves, and we specify $p(o|w)$ to be the way in which the organism receives noisy and partial observations.

\subsection{Attempts So Far}
\label{sec:1}

The first instance of such an objective function incorporating sensors and actuators is perhaps a paper by Still \cite{still2009information}. She imagined that an organism sees observations $o_t$ at time $t$, converts past actions and observations to sensory state $s_t$, and takes action $a_t$ right after based also on that history. The history of observations and actions is labeled $h_t$ and the future of observations is labeled $z_t$. She imagines that $h_t$ is used to inform both $s_t$ and $a_t$ separately. Still suggests that one should try to maximize $I[s,a;z]-\lambda I[s;h] -\mu I[a;h]$ where $\lambda,~\mu$ are Lagrange multipliers and time indices have been dropped for easier-to-read notation. In this objective, $I[s,a;z]$ is the mutual information between the sensory state and the action relative to the future of observations; $I[s;h]$ is the mutual information between sensory state and history; and $I[a;h]$ is the mutual information between action and history. In Ref. \cite{still2009information}, Still found optimal sensors to be sensorimotor causal states (described in Sec. \ref{sec:3}) in the limit that $\lambda\rightarrow 0$ and also identified optimal action policies in the limit that $\mu\rightarrow 0$.

The first term in this objective is interesting, but maximizing this term usually leads to large periodic loops when $\lambda,~\mu$ are near enough to $0$. (Large periodic loops have a high mutual information between past and future.) That is unfortunately a limit of interest for higher-level organisms that can pick up the aforementioned sensorimotor causal states. Although some work \cite{bialek2001predictability} claims that these high predictive information processes correspond to processes that learn underlying parameters of the environment model, that is only true in a nonergodic case \cite{crutchfield2015signatures}. It may be possible in certain environments to see something more complex \cite{marzen2016statistical}. For lower-level organisms, the limit $\lambda,~\mu\rightarrow\infty$ is of greater interest, but that leads to sensory states and actions that depend not at all on the history and are instead biased coin flips, by simulations not shown here. A quick theoretical argument suggests that should be the case-- $I[s;h],~I[a;h]$ can both be set to $0$ if $s,~a$ have no dependence on $h$, and thus the objective function is maximized by doing so.

The next instance of such an instantiation that is information-theoretic comes identically from Ref. \cite{lai2023human} and Ref. \cite{malloy2021capacity}. Here, the information-theoretic term $I[s,a;z]$ is replaced by the usual reinforcement learning term $V_{\pi}$, the sum total of discounted rewards. Rewards depend on the underlying Markov state of the environment $w_t$, so that $V_{\pi} = \sum_t \gamma^t r(w_t,a_t)$ where $\gamma$ is a discount factor, $r$ the usual reward function \cite{sutton2018reinforcement}, and $w_t$ and $a_t$ the world state and actions at time $t$. It is straightforward to generalize to continuous-time by introducing an integral. There is no cost for complicated sensory states $s$, unlike in Ref. \cite{still2009information}. There is only a cost on transmitting information from sensory state to actions $I[s;a]$, the mutual information between sensory state $s$ and action $a$. As a result, the objective function reads $V_{\pi}-\beta I[s;a]$. Note that here, $s$ is used to inform the action $a$ rather than the entire history $h$ being used to inform the action. This rings more true to neuroscience, as we describe in the next section.

The work from Ref. \cite{arumugam2024bayesian} looks similar in spirit to the second of these two instantiations, but there, the rate constraint is included for a completely different reason. It encourages exploration in complex environments. The work in Ref. \cite{NEURIPS2022_8bb5f663} for language also looks similar, but there, two terms exist that encourage understanding the environment-- one that resembles an information bottleneck term that maximizes mutual information with the relevant variable, and one that maximizes utility, while one term remains to penalize understanding the environment.

Our work is maybe closest to Ref. \cite{tishby2010information,van2012informational} which includes both the rate constraint on communicating sensations to actions and also a sensory variable that is recursively updated as in the Recursive Information Bottleneck (RIB) \cite{still2014information}. The RIB constraint allows for a large value of resources when the brain is large and does not even take in any information about the stimulus, and hence seems less of a notion of memory than the information-theoretic quantity we use here. However, in their eventual algorithm, sensor and world states are collapsed into one, which we avoid as the sensory system is a bottleneck for information about the world.


\subsection{The New Objective}
\label{sec:2}

We must carefully decide which terms to include in the final objective function describing an organism trying to navigate a sensorimotor feedback loop. Altogether, we would like an objective function that naturally balances exploration and exploitation, meaning that an organism should explore its environment naturally before exploiting the information it has obtained to survive; and we would like an objective function that includes as many resource constraints as possible. Exploitation naturally requires exploration, since to exploit means that one has sampled the environment enough to know which action is best, as can be seen when considering a simple multi-armed bandit. Potentially an objective function could start with more emphasis on exploration to encourage better exploitation later. A simple combination of the objective functions that exist so far as mentioned in Sec. \ref{sec:1} yields:
\begin{equation}
\mathcal{L} = V_{\pi} - \beta I[s;a] -\lambda I[h;s]
\end{equation}
where $\beta,~\lambda$ are constants.
This is really the unconstrained version of a constrained objective function:
\begin{equation}
R(MI_{s,a},MI_{h,s}) = \max_{I[s;a]\leq MI_{s,a},I[h;s]\leq MI_{h,s}} V_{\pi}
\end{equation}
so that $\beta,~\lambda$ are Lagrange multipliers and $MI_{s,a}$ and $MI_{h,s}$ are adjustable constants.

It is possible that this unconstrained objective function is itself more fundamental than the constrained version of the objective function, with reward being offset by costs. For example, the reward function is essentially equivalent to energy-gathering, while the two resource constraints linearly combined relate to energy expenditure. If so, the Lagrange multipliers may attain physical meaning that translates rates into energies, e.g. temperatures multiplied by the Boltzmann constant.

With the constrained objective function, we define the reward-rate manifold, in which $MI_{h,s}$ is on the $x$-axis, $MI_{s,a}$ is on the $y$-axis, and $V_{\pi}$ on the $z$-axis. The manifold separates achievable combinations of information-theoretic rates $I[h;s],~I[s;a]$ and rewards $V_{\pi}$ and unachievable combinations, as in rate-distortion theory \cite{berger} and predictive rate-distortion theory \cite{marzen2016predictive}. In other words, the reward-rate manifold defines a Pareto front. 

First, we discuss the term that allows the organism to accumulate reward. The term $V_{\pi}$ naturally implies that we must both explore and exploit: to reap rewards, one must survey all available options (within reason) and choose the best one rather than merely sticking with the first good option that comes around. However, much effort has been spent in reinforcement learning trying to add additional terms or alter action policies so that a better balance of exploration and exploitation is achieved, e.g. as in Ref. \cite{burnetas1996optimal}.

Next, we discuss the information-theoretic resource term that suggests the organism should aim for a simpler actuator. We must convey the sensory state $s$ to find the action policy $a$ using the conditional probability $\pi(a|s)$ that signifies the action policy \cite{sutton2018reinforcement}-- the actuator $a$ does not have direct access to histories $h$-- and so $I[s;a]$ is the appropriate term, as identified by Refs. \cite{lai2023human, malloy2021capacity}.

Finally, we discuss the information-theoretic resource term that suggests the organism should aim for a simpler sensory layer \cite{palmer2015predictive}. If we think about the human brain, observations from the retina $o$ must combine with efference copies $a$ at the primary visual cortex V1 to give us a sensory state $s$ that can be used to determine actions. Mathematically, there is some input-dependent dynamical system that takes in information from the efference copy and the observations and turns it into something that is not quite the history $h$ written down by Still, but has information going back to the beginning of when the organism has opened its eyes. Hence we are perhaps somewhat justified in replacing this variable by $h$. This information must be communicated to the next layer in the brain, justifying $I[h;s]$ as the next resource constraint. 

Evolution is not likely to directly work on this objective function, but might be subject to resource constraints that force it to essentially maximize this objective function. Essentially, the resource constraints that evolution operates on might look more like material constraints \cite{chklovskii2004maps} or energy constraints \cite{hasenstaub2010metabolic,mehta2012energetic}, both which lead to mutual informations as the natural stand-in using results from information theory or nonequilibrium thermodynamics. See App. \ref{app:1}.

\section{An Algorithm To Calculate Using Sensorimotor Causal States}
\label{sec:3}

Sensorimotor causal states as defined in Ref. \cite{still2009information} are usually also belief states of the POMDP \cite{kaelbling1998planning}. Belief states are the probability distribution over the underlying Markov state of the environment (or more technically, of the POMDP) $w$ given the history $h$, and one uses these to ``solve'' the POMDP-- to determine one's action policy \cite{kaelbling1998planning,doshi2013bayesian}.

These sensorimotor causal states come from a coarse-graining relationship, as in Ref. \cite{still2009information,shalizi2001computational}. Take histories $h$ and consider two histories $h,~h'$ equivalent if $P(w|h) = P(w|h')$. Note the difference from Ref. \cite{still2009information}-- we have replaced future observations with the underlying Markov state of the POMDP. The best guide to the future of the observations is the underlying Markov state of the environment $w$. This is unobtainable directly, so in any real algorithm to ascertain sensorimotor causal states, one might use the future of observations instead. Regardless, the clusters of histories are labeled $\sigma$, sensorimotor causal states, and the sensorimotor causal state to which history $h$ belongs is given by $\epsilon^+(h)$. We define sensorimotor causal states in this modified way so that the proof of the main theorem in this paper is clear; as an added benefit, these modified sensorimotor causal states are now \emph{exactly} the belief states.

With this definition in hand, we introduce our main theorem and proof that simplifies calculation of the reward-rate manifold.

\begin{theorem}
The objective function from the previous section was $V_{\pi} - \beta I[s;a] - \lambda I[h;s]$. We can replace histories $h$ with sensorimotor causal states $\sigma$ if we wish to find statistics of good sensors \cite{berger} or to calculate the reward-rate manifold.
\end{theorem}

To prove this, note that there is no change to $V_{\pi}$ or $I[s;a]$ if sensory states $p(s|h)$ are recoded as $p(s|\sigma=\epsilon^+(h))$, similar to what is true in Ref. \cite{marzen2016predictive}. And, as in Ref. \cite{marzen2016predictive}, $I[s;h] = I[s;\sigma] + I[s;h|\sigma]$ only decreases with this recoding to $I[s;\sigma]$ since $I[s;h|\sigma]\geq 0$. The objective function therefore benefits from this recoding. As a result, as expected, it is optimal to pick up sensorimotor causal states using the recurrent neural network that governs the sensory layer in biological organisms.

The new insight into sensory states is that they should pick up nothing else, however lossy; and that the objective function can be rewritten with histories $h$ replaced with sensorimotor causal states $\sigma$.

Importantly, the obtained sensor $p(s|h)$ and actuator $\pi(a|s)$ from maximizing this objective might not be good sensors or actuators themselves by the original material constraints \cite{berger}. This is a common misconception for practitioners of the information bottleneck method, as this point is not stressed by the seminal work in Ref. \cite{tishby2000information}. (The information bottleneck method is a rate-distortion method with an informational distortion.) The soft clusters obtained by the Blahut-Arimoto algorithm and generalized Blahut-Arimoto algorithm are often bad lossy compressors due to the difference between $H[a]$ and $I[a;s]$, where $I[a;s]$ is typically considered to be the resource constraint. But it is sadly the case that $H[a]$ and not $I[s;a]$ mirrors the expected length of the coding of the action sequence, and $H[a]$ is almost always larger, and maybe much larger. Also, as a single-symbol compression scheme, the codes revealed by these iterative algorithms are not usually optimal, except for special cases of the distortion measure \cite{moffett2021code}. This is true even when $H[a]$ replaces $I[a;s]$ in the objective function \cite{marzen2024comment}. If several symbols are used, as is more typical for good lossy compression schemes, then the statistics of a good lossy compression scheme will mirror the soft clusters obtained by the IB method \cite{berger}.

We now specialize to the case of no discounting $\gamma=1$, in which case $V_{\pi}$ turns into a sum of rewards, for ease, with anticipated extensions later. For a POMDP, one can define a reward function on belief states $\sigma$ and actions $a$ from the underlying reward function on underlying Markov states of the environment $w$ and actions $a$ \cite{kaelbling1998planning}, but we avoid this step. (It is not necessary for calculating the reward-rate manifold for the experiments we plan to do in the future.)
Under a stationarity condition, $V_{\pi}$ turns into $T\langle r(w,a)\rangle_{p(w,a)}$, where $T$ is the total number of time steps in the organism's life, and $\langle \cdot \rangle$ is an expectation value, replacing what is often labeled as $E[\cdot]$. We can ignore the additional factor of $T$ by rescaling $\beta,~\lambda$.

In this case, from Appendix \ref{app:2}, we can calculate the reward-rate manifold by using the iterative algorithm which updates $\pi_n(a|s)$ and $p_{n}(s|\sigma)$ as in the usual information bottleneck algorithm \cite{tishby2000information}:
\begin{widetext}
\begin{eqnarray}
\pi_{n+1}(a|s) &=& \pi_n(a) \frac{\exp \left(\frac{1}{\beta}\sum_{\sigma,w} p_n(\sigma|s) p_n(w|\sigma) r(w,a)\right)}{Z_{\beta,n}(s)} \label{eq:3}
\end{eqnarray}
where $Z_{\beta,n}(a)$ is a partition function or normalization factor, similar to Refs. \cite{tishby2000information,still2009information}, so that
\begin{eqnarray}
Z_{\beta,n}(a) &=& \sum_a \pi_n(a) \exp\left(\frac{1}{\beta} \sum_{\sigma,w} p_n(\sigma|s) p_n(w|\sigma) r(w,a)\right).
\end{eqnarray}
Similar manipulations for $p(s|\sigma)$ gives
\begin{equation}
p_{n+1}(s|\sigma) = \frac{p_n(s) \exp\left(\frac{1}{\lambda} \sum_{a,w} \pi_n(a|s) p_n(w|\sigma) r(w,a)\right)}{Z_{\lambda,n}(\sigma)} \label{eq:4}
\end{equation}
where $Z_{\lambda,n}(\sigma)$ is again a partition function or normalization factor,
\begin{eqnarray}
Z_{\lambda,n}(\sigma) &=& \sum_s p_n(s) \exp\left(\frac{1}{\lambda} \sum_{a,w} \pi_n(a|s) p_n(w|\sigma) r(w,a)\right).
\label{eq:6}
\end{eqnarray}
As the action policy and sensory apparatus change with iteration, so do the sensorimotor causal states and their relationship to the underlying world states. We use a combination of the algorithms in Refs. \cite{crutchfield2016exact,marzen2017nearly} to tackle this problem, as described in Appendix \ref{app:2} and in Algorithm \ref{alg:cap1} from $p(o|w)$, $p(w_{t+1}|w_t,a_t)$, $p_n(s|\sigma)$, and $\pi_n(s|a)$, where there is an adjustable parameter $N$ governing the length of the observation sequence used to estimate $p_{n+1}(w|\sigma)$ and $p_{n+1}(\sigma)$. Interestingly, this aspect of the algorithm is missing in Ref. \cite{still2009information}'s variational treatment, since that treatment does not take into account the fact that her $P(a|h)$ affects her $P(z)$ in unanticipated ways due to sensorimotor feedback, for example-- the action policy affects all future observations not just via a marginalization over one time step, but all time steps.

\begin{algorithm}[H]
\caption{The sensorimotor causal states algorithm to find the reward-rate manifold }\label{alg:cap1}
\begin{algorithmic}
\State Input world characteristics $p(o|w)$ and $p(w_{t+1}|a_t,w_t)$, and organism's relationship to the environment $r(w,a)$.
\While{$\beta$, $\lambda$ run through a list of possible $\beta$'s, $\lambda$'s that trace out the manifold}
\State Initialize $p(s|\sigma)$, $p(a|s)$.
\State Calculate the corresponding $p(w)$ and then use the mixed state presentation to find the causal states. \Comment{The length of observation sequences $N$ and the resolution of the simplex $\epsilon$ are hyperparameters. $\epsilon$ should be as small as possible and $N$ as large as possible without sacrificing computational efficiency for consistency.}
\State From the causal states, find $p(s|\sigma)$ by averaging the $p(s|h)$ for those $h$ in the same sensorimotor causal state.
\State Run Eqs. \ref{eq:3}-\ref{eq:6} to convergence.
\State Collect $I[s;a]$ and $I[s;\sigma]$ and $\langle r(w,a)\rangle$ for that $\beta$, $\lambda$.
\EndWhile
\State Parametrically plot the reward-rate manifold.
\end{algorithmic}
\end{algorithm}

\end{widetext}
As $\lambda,~\beta$ change from $0$ to $\infty$, we can trace out the entire two-dimensional reward-rate manifold. Because the objective function is convex in the sensor description $p(s|\sigma)$ and actuator description $\pi(a|s)$, this generalized Blahut-Arimoto algorithm will converge roughly to the global optimum as $n\rightarrow\infty$, with the caveat that $N$, $\epsilon$ controls the quality of convergence. We wish to make $N$ as large as possible and the coarse-graining of the simplex $\epsilon$ as small as possible, but also require compute efficiency.

\begin{theorem}
The objective function, in the limit $\lambda\rightarrow 0$, finds that the sensor states should be sensorimotor causal states, and in the limit $\beta\rightarrow 0$, finds that the action policy should be deterministic.
\end{theorem}

As in Ref. \cite{still2009information}, in the limit that $\lambda\rightarrow 0$, we find that $s$ recovers exactly the sensorimotor causal states $\sigma$ and in the limit that $\beta\rightarrow 0$, we find a deterministic action policy. To see this, we can simply stare at the objective function and note that as these Lagrange multipliers tend to $0$, our goal is to maximize reward and we do not care about the rates, which is accomplished when you store as much information about the environment as possible in your sensor and have a one-to-one mapping from sensory states to actions. This lossless limit is likely nearby to what humans or very complex organisms experience. Then, in the limit that $\beta,~\lambda$ are large, we find that the sensor picks up no information about the causal state and that the actuator is completely stochastic and does not depend on sensor state, simply from glancing at the importance of the mutual informations in the objective function in this limit. This lossy limit is likely close to what small fish or other simple organisms experience. However, again, the goal here is not necessarily to find sensors or actuators-- though by conjecture \emph{statistics} of good ones can be obtained from this algorithm \cite{berger}-- but to calculate a reward-rate manifold so as to benchmark how well biological and artificial agents reap reward under resource constraints in POMDP environments.

Note that before this theorem, operating on long histories to calculate the reward-rate manifold would encounter two curses of dimensionality based on the length of the history. We have replaced histories with sensorimotor causal states, bypassing one curse of dimensionality \cite{barnett2015computational}, as in Ref. \cite{marzen2016predictive}. Still, a curse of dimensionality is encountered from the algorithm in Ref. \cite{marzen2017nearly}. One can calculate, in theory, the reward-rate manifold from an algorithm like that of Algorithm \ref{alg:cap1}, with more algorithms to come in future work.


\section{An Example Reward-Rate Manifold}
\label{sec:4}

Interestingly, $\epsilon$ and $L$ need be so small and large for simple POMDPs that we readily encounter a curse of dimensionality with the algorithm as written. Improvements will need to be made before it can be used with confidence. But that does not mean we cannot use the variational equations derived above to produce an example reward-rate manifold.

We start with what the environment provides, decided arbitrarily for this example to be:
\begin{eqnarray}
w_{t+1} &=& w_t-a_t+\eta_{w_t} \label{eq:1} \\
o_t &=& w_t + .01 \eta_{o_t} \label{eq:2} \\
r(w_t,a_t) &=& -\langle (a_t-w_t)^2\rangle \label{eq:3}.
\end{eqnarray}
Here, $\eta_{w_t},~\eta_{o_t}$ are zero-mean, unit-variance Gaussian noise. The first of these equations, Eq. \ref{eq:1}, describes how the environment evolves with sensorimotor feedback, and the second, Eq. \ref{eq:2}, describes how the observation is the world state corrupted with a tiny bit of noise. Because the noise is so tiny, roughly speaking, $\sigma$ is $w$. Thus, the objective function becomes
\begin{equation}
\mathcal{L} = -\langle (a-w)^2\rangle - \beta I[s;a] - \lambda I[w;s].
\end{equation}
The last environmental setup equation, Eq. \ref{eq:3}, is a reward function that wishes for $a_t$ and $w_t$ to be as different as possible.

It turns out that linear dynamical systems with Gaussian noise the entire way through-- for how $s$ relates to $o$ and for how $a$ relates to $s$-- solve the generalized Blahut-Arimoto equations in Sec. \ref{sec:3}. Therefore, we can write that to solve the objective function,
\begin{eqnarray}
s_t &=& m_{s,o} o_t + \sigma_{s,o}\eta_{s_t} \\
a_t &=& m_{a,s} s_t + \sigma_{a,s} \eta_{a_t},
\end{eqnarray}
where $\eta_{s_t},~\eta_{a_t}$ are again zero-mean, unit-variance Gaussian noise. We avoid additive constants to avoid a trivial solution in which the reward is maximized by simply making additive constants as large as possible, though we leave a full discussion of this phenomenon for later work.

\begin{figure}
\centering
\includegraphics[width=0.45\textwidth]{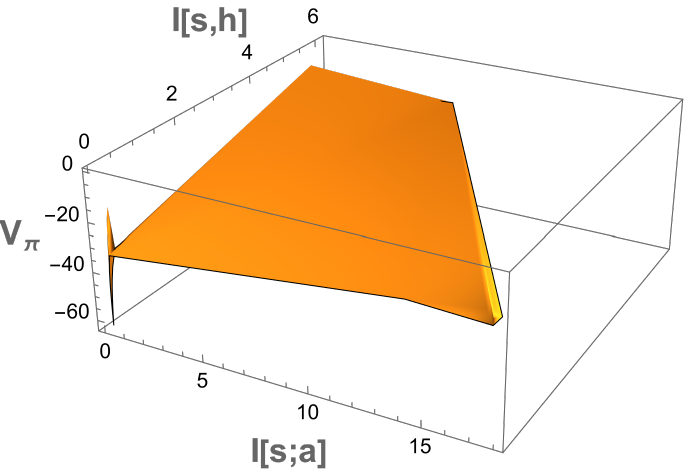}
\caption{For the environment described in Sec. \ref{sec:4}, an approximate reward-rate manifold constructed using Appendix \ref{app:4}'s equations leading to numerical maximization of the objective function in Sec. \ref{sec:3} in Mathematica for several $\beta,\lambda$ between $0$ and $1000$. As expected, we see a surface that approximately (within numerical error) monotonically increases the reward as both rates increase. The surface resembles the strange information curve behavior of the Even Process in Ref. \cite{marzen2016predictive}.}
\label{fig:1}
\end{figure}

We must solve for $m_{s,o},~m_{a,s},~\sigma_{s,o},~\sigma_{a,s}$ to maximize $\mathcal{L}$, which we do numerically. We just need slightly simpler expressions for all the mutual informations and the expected reward, which we find in Appendix \ref{app:4}. The resultant approximate reward-rate manifold is shown in Fig. \ref{fig:1}.

\section{For Which Organisms Might Resource-Rational Reinforcement Learning Fail?}

From Ref. \cite{celani2010bacterial}, it is clear that bacteria are maximiners that choose a strategy that works best for the worst-case scenario rather than operating on a discounted sum of rewards. Surprisingly, this means that bacteria do not actually seek to climb chemoattractant gradients unlike as stated in many references, including even Ref. \cite{mattingly2021escherichia}. This is despite the fact that climbing chemoattractant gradients would be an obvious strategy for thriving. Instead, bacteria live in environments in which chemoattractant gradients leave about as quickly as they come, as every other bacterium also tries to climb and eat the chemoattractant. To operate well, they assume a worst-case environment, and do the best job they can given that pessimistic assumption.

That is only one data point, but it is enough to make us pause. We have ignored these lower-level organisms for the purpose of this paper so far, but we shouldn't. They are an important part of biophysics. We understand them far better than we understand humans, simply due to the lower complexity of the bacterium. In this section, we argue that there is likely a phase transition as organisms increase in complexity from the maximin behavior of the bacterium to the complicated reinforcement learning strategies of humans. 

In particular, we conjecture that for organisms of low enough complexity, you see such maximin behavior only because the organism lacks a theory of mind-- an inability to understand environments that are actually other reinforcement learning agents with desires of their own. When an organism can understand other organism's desires, they have an ability to exploit complicated environments that have agency simply because they are partly composed of other organisms \cite{seifert2024reinforcement}, and therefore can operate by maximizing a discounted sum of rewards. Otherwise, we would argue, the organism in question does its best job by assuming a worst-case scenario, or assuming an adversarial environment, like the bacterium faces. Therefore, there could be a phase transition in Marr's computational level objective as organisms increase in complexity from those that lack of theory of mind to those that possess one, based simply on the size of the corresponding brain region. The theory of mind may even be quite simple and quite implicit, as C. elegans may have enough of a theory of mind to be better described by the resource-rational reinforcement learning described so far. The question we must essentially ask to see if the organism falls on one or the other side of the phase transition is: does the organism have anything resembling mirror neurons?

It could also be the case that higher-level organisms look as though they are maximizing something like a discounted sum of rewards, but that actually, they have found the best behavior for the worst-case environment. Perhaps we have only tested higher-level organisms in environments where the maximin behavior is surprisingly close to the policies that you would get from standard reinforcement learning.

How would we know the difference between a phase transition to resource-rational reinforcement learning and maximin behavior that looks like resource-rational reinforcement learning? There is a simple way to test whether or not a higher-level organism is accomplishing resource-rational reinforcement learning or simply using the maximin action policy, and it involves finding an environment in which these two are very different. To show that this is possible, we now-- for the random environment described later-- place the maximin behavior relative to the reward-rate manifold. To find the maximin behavior, we allow for the organism's sensory system to store the sensorimotor causal state just so that we can get some insight into the maximin action policy relative to the reward-rate manifold. We then look for an action policy $\pi(a|\sigma)$ that solves
\begin{equation}
\pi_{minimax}(a|\sigma) = \arg\max_{\pi(a|\sigma)}\min_{p(w|\sigma)} \langle r(w,a)\rangle.
\end{equation}
As the environment changes, $p(w|\sigma)$ morphs, and so we assume for these lower-level organisms that they are assuming pessimistically an environment that has the worst possible $p(w|\sigma)$ imaginable. It may not be possible to achieve this particular worst-case scenario given $p(o_t|w_t)$ and $p(w_{t+1}|a_t,w_t)$ and yet we assume this to make progress. Note that it is probably unreasonable to assume a lower-level organism can store sensorimotor causal states rather than lossy sensorimotor causal states, but this optimistically gives us our best shot at reaching the reward-rate manifold to see whether or not we can spot the difference between the two objectives even in simple random environments. See Appendix \ref{app:3} for an approximate solution to this maximin objective based simply on multivariable calculus. See Algorithm \ref{alg:cap2}. One simply calculates the reward and rates of the maximin strategy and compares to the relevant point for the iterative algorithm-- one where $\lambda,\beta\rightarrow 0$.

\begin{algorithm}[H]
\caption{The approximate maximin solution}\label{alg:cap2}
\begin{algorithmic}
\State Input $r(w,a)$.
\State Use Eqs. \ref{eq:C10}-\ref{eq:C13} to calculate an approximate maximin solution.
\end{algorithmic}
\end{algorithm}


This, incidentally, illustrates how one can test if an organism is a resource-rational reinforcement learner. One simply measures the organism's behavior and sensory states using some sort of neural or other readout and calculates the relevant rates, $I[s;h]$ and $I[s;a]$, and reward, $\langle r\rangle$. Then this point is compared to the reward-rate manifold, finding like rewards and comparing rates or finding like rates and comparing rewards. Like in rate-distortion theory, if this point is close to the reward-rate manifold, we deem the organism a nearly optimal resource-rational reinforcement learning, as in Refs. \cite{palmer2015predictive,lamberti2023prediction} for resource-rational prediction. If this point is not close, perhaps relative to a null model of some kind, then the organism is not a resource-rational reinforcement learner by the end of the experiment.

In general, lower-level organisms are unlikely to be able to pick up on the full sensorimotor causal state. (We simply assumed they could for illustrative purposes.) Perhaps instead, we can view lower-level organisms as having resource constraints that force $p(s|h)$ to fall into a certain parameterized family $\mathcal{F}$ and that force $\pi(a|s)$ to fall into another parameterized family $\mathcal{G}$. A resource-rational maximiner, then, would take the form
\begin{widetext}
\begin{equation}
\pi_{maximin}(a|s),~p_{maximin}(s|h) = \arg\max_{\pi(a|s)\in\mathcal{G},~p(s|h)\in\mathcal{F}} \min_{p(w_{t+1}|a_t,w_t)} \langle r(w,a)\rangle
\end{equation}
\end{widetext}
with $\mathcal{F},~\mathcal{G}$ to depend on mechanistic details of the organism. We leave this as an intriguing proposal for what a lower-level organism might be trying to do and also leaving experimentalists to test whether or not higher-level organisms are resource-rational maximiners instead of resource-rational reinforcement learners.

\section{Conclusion}

In this manuscript, we have proposed a new computational-level objective function for theoretical biology and theoretical neuroscience that combines: reinforcement learning \cite{sutton2018reinforcement}, the study of learning with feedback via rewards; rate-distortion theory, a branch of information theory \cite{berger,tishby2000information} that deals with compressing signals to retain relevant information; and computational mechanics, the study of minimal sufficient statistics of prediction also known as causal states \cite{shalizi2001computational,still2009information}. We have highlighted why this proposal is likely only an approximation, but is likely to be an interesting one, and proposed a new algorithm for evaluating it to obtain the newly-coined ``reward-rate manifold''.

The reward-rate manifold is like a rate-distortion function, but in a system where there is both feedback and memory (an underexplored area in information theory) and with one additional rate so that not just the sensor is considered, but the actuator too. Due to the difficulty of analyzing memoryful communication channels with feedback and memoryful input in information theory, we have merely conjectured that this reward-rate manifold might provide a guide to how biological organisms function, in the same way that the predictive rate-distortion function provided insight into the salamander retina \cite{palmer2015predictive} and cultured neurons \cite{lamberti2023prediction}.

It is important to stress that biological organisms are likely not operating directly on this objective function. Rather, they are naturally subject to resource constraints that lead to them naturally maximizing this objective function. Nor are the sensors and actuators revealed by this objective function likely to be the actual sensors and actuators used-- famously, the sensors and actuators that are revealed only provide statistics that describe the true sensors and actuators that do well on the objective function \cite{berger}.

In order to calculate this reward-rate manifold, it will usually be necessary to use the sensorimotor causal states first proposed in Ref. \cite{still2009information}, although the algorithm implemented here in Appendix \ref{app:2} still encounters a curse of dimensionality, unfortunately. One might reasonably ask why the organism should have access to the sensorimotor causal states. Rather, the organism is likely trying to infer sensorimotor causal states using some algorithm that we have not yet determined \cite{doshi2013bayesian,barnett2015computational}. As in Refs. \cite{palmer2015predictive,pnasnexus,humanbehavior}, we envision a raft of experiments that involve the experimentalist knowing the environmental statistics with which the organisms are probed and using their knowledge of sensorimotor causal states to calculate the reward-rate manifold, calculate the reward and rates of the organism from behavioral and neural data, and then place the organism's operation relative to the reward-rate manifold as is common in rate-distortion theory \cite{berger}. This will enable a stringent test of whether or not the organism really is maximizing expected reward subject to information-theoretic rate constraints, as we have done here with approximations to the maximiner (bacteria-like) strategy.

At this point, it is crucial to note that the iterative algorithms used to find the reward-rate manifold and the brute force algorithm used to find the maximiner strategy should be improved upon. The reward-rate manifold's iterative algorithm derived in Appendix \ref{app:2} is elegant enough when considering updates for the sensor and the actuator, but due to feedback, it is complicated to find the new sensorimotor causal states. For that, we used the algorithm in Ref. \cite{marzen2017nearly}, which encounters a curse of dimensionality. This is hard to avoid, as typically, there are an uncountable infinity of sensorimotor causal states, and we merely approximate them with a partition of the belief state space. We envision improvements might come from a variational algorithm using neural networks like that of Ref. \cite{hahn2019estimating} or like that of Refs. \cite{alemi2020variational,alemi2016deep}, or potentially using a Gaussian Information Bottleneck-like algorithm as in Ref. \cite{chechik2003information}. A Gaussian Information Bottleneck-like approach, based on the iterative equations proving that a self-consistent solution was Gaussian, was used in Section \ref{sec:4} to find an approximate example reward-rate manifold. In Appendix \ref{app:3}, the maximiner strategy assumed that an optimal sensorimotor causal state distribution could be obtained, but this is likely not true in general, and while the foundations for finding the correct maximiner strategy are in this paper, the algorithm is not.

Future work will center on calculating this reward-rate manifold for various environments and placing organism brain recordings and behavioral assays relative to the reward-rate manifold.

This is likely only a first approximation to the true computational-level objective. The most important objection we have comes from what we consider ``memory'' to be, which is, at present, a nonlinear correlation coefficient between stimulus past and sensory brain state \cite{kinney2014equitability}. This is correlated with working memory in one experiment to date \cite{humanbehavior}, but this is just one experiment. Plus, memory is quite complicated and extends far beyond working memory \cite{sridhar2023cognitive}.

In fact, it is not even clear that memory is the right resource to look at. What about a thermodynamic constraint like entropy production rate, which is lower-bounded \cite{still2012thermodynamics} (sometimes loosely \cite{marzen2020prediction}) by nonpredictive information rate? Or are energetics irrelevant as resource constraints for a system of this size and processing power, despite some beautiful experiments on lower-level situations that are more amenable to mechanistic analysis \cite{hasenstaub2010metabolic}?
Future efforts might focus on including time, as much effort has been spent understanding the speed-accuracy-energy tradeoff in nonequilibrium thermodynamics \cite{lan2012energy}, or notions of processing power and computability. In other work, minimum description length might even replace mutual informations \cite{moskovitz2024understanding}. This reward-rate manifold is just the start to what might appear, as more ``rates'' are added that may not even be mutual informations. The point of this paper is to propose the idea of a reward-rate manifold, which allows direct testing of all of these normative principles for brain and behavior of organisms simply by plotting where the neural recordings and behavior lie relative to a reward-rate manifold.

This proposal does not solve at all the algorithmic or mechanistic level, although ideas about the mechanistic level have informed the very foundations of this computational-level objective via resource constraints. However, this computational-level objective and the algorithm used to find its associated sensors and actuators cannot be compared to the algorithmic and mechanistic levels, for interesting reasons rooted in rate-distortion theory \cite{berger}. Thus, those algorithmic and mechanistic details are left to methods such as maximum likelihood determination of the true sensory and actuator strategies \cite{uppal2020inferring,daw2011trial}. Still, we hope that this contribution allows for the development of a research program that will finally unfurl the computational level of theoretical biology and theoretical neuroscience.

And really, the aim is quite ambitious, as we wish to describe all organisms-- not just humans-- with a theory of mind by one objective function that is altered to the specifics of the organism's situation only by a change in the POMDP and the Lagrange multipliers for the resource constraints (or equivalently, the level of resources themselves). There may be variation in a population as to how close to the Pareto front organisms are or their individual level of allowed resources for a particular computation as in prior work \cite{lamberti2023prediction, humanbehavior, lai2024human} with a tendency to dot the Pareto front \cite{tkavcik2025information}, but we expect that humans as a group have a strikingly different level of allowed resources than mice or fish in general that will depend on how much the organism \emph{cares} about the specific task being tested.

Looking to the future, we can of course not rule the possibility that some objective functions might explain biological data well \cite{lai2023human,pnasnexus,palmer2015predictive,humanbehavior} but be later superseded, as one instantiation of the efficient coding hypothesis \cite{laughlin1981simple} was later replaced by another \cite{park2017bayesian}. But we do hope that this objective function and others of its ilk provide a start towards testing if organisms are ``good enough'' or actually resource-rational decision makers.

\begin{acknowledgments}
I would like to thank Dmitri Chklovskii and Rainer Engelken for inspiring conversations, and I would like to thank Lav Varshney, Artemy Kolchinsky, and Joshua Shaevitz for very helpful comments. Thank you also to anonymous referees for comments.
\end{acknowledgments}

\begin{widetext}
\appendix

\section{Reasoning for Mutual Informations From the Rate-Distortion Theorem}
\label{app:1}

Before we describe the resource constraints for this POMDP, let us describe the rate-distortion theorem \cite{berger}. It will justify why material constraints can likely be replaced by mutual informations.

In the classic rate-distortion setup, one sends a sequence of $n$ letters $x_{0:n}$ to an encoder that chooses one of $M$ words for those $n$ letters and then sends that word to a decoder which produces a guess as to what those letters were, $\hat{x}_{0:n}$. The material constraint is actually $\log M/n$, not a mutual information. This corresponds to a more intuitive notion of resource constraints in the biological sense-- number of molecules or number of neurons, normalized by ``blocklength'' $n$. Some distortion measure is defined, $d(x,\hat{x})$, which could be generalized to a distortion of the entire block $x_{0:n}$ relative to $\hat{x}_{0:n}$ rather than letter-by-letter, also called an $n$-letter extension. There are some rates $\log M/n$ and distortions $\sum d(x_i,\hat{x}_i)/n$ that are achievable and some that are unachievable given any combination of encoder and decoder. A theorem shows that the curve separating achievable from unachievable is given by replacing the rate $\log M/n$ with a mutual information $I[X;\hat{X}]$ and the average distortion with an expected distortion if all is memoryless. This curve is accurate in the limit that blocklength $n$ goes to infinity. Otherwise, the rate-distortion curve that separates achievable from unachievable is given by $R_n(D)$ rather than $R(D)$, and $R_n(D)$ is horribly difficult to calculate \cite{berger}, but see Ref. \cite{kostina2012fixed}. In essence, what we will try to argue is that biological organisms operate in the limit of very large $n$ sometimes, and so it is okay to use mutual informations to calculate the ``reward-rate manifold''-- the two-dimensional manifold that separates allowable from unallowable combinations of the two rates to be discussed and the reward $V_{\pi}$. Otherwise, $R_{n}(D)$ places an upper bound on $R(D)$, and since the reward is the flip of the distortion, the corresponding logic is that $R_n(MI_{s,a},MI_{h,s})$ places a lower bound on $R(MI_{s,a},MI_{h,s})$.

The key material constraint that we wish to think about is the number of neurons, either in the sensory layer or in the actuator layer. If there is a combinatorial code, then the number of words $M$ is equivalent to $2^{num}$ where $num$ is the number of neurons. A resource constraint that is reasonable is therefore $\log M$. This must be modulated by a blocklength-- some sense of timescales. The NMJ (neuromuscular junction, or actuator layer) is thought to operate by a rate code, while the sensory layers are thought to operate on sub-millisecond timescales \cite{nemenman2008neural} and the environment is thought to operate on extremely large timescales given that naturalistic video is described by power laws. Given all this, the effective blocklength for the actuators is likely to be very high, so that $I[s;a]$ is justified. Then again, blocklengths are costly for reward reasons \cite{sawaya2023framework}, but we leave the question of why this mutual information constraint appears to explain biological data to some extent in reinforcement learning experiments \cite{lai2024human} as an anomaly to be figured out by future practitioners. And, $I[h;s]$ provides us with a lower bound on the reward-rate function and appears to correlate with working memory in at least one study \cite{humanbehavior} and that explains biological data in other studies \cite{palmer2015predictive,pnasnexus}.

A complication exists with what seems to be an exquisite theoretical justification from information theory: the environment is memoryful, and so are the sensors and actuators. The rate-distortion theorem does extend to stationary, ergodic processes. However, memoryful processes have much harder-to-calculate objectives because the entire sequence of inputs and outputs is considered in the rate constraint \cite{berger}, though see Ref. \cite{arnold2006simulation} for algorithms to compute the rate. As a result, we replace material constraints with mutual informations by conjecture as an approximation to what is likely true.

In a thermodynamic direction, Landauer-like bounds suggest that mutual informations might lower-bound dissipated work \cite{sagawa2009minimal, still2012thermodynamics}.

Even if none of these information-theoretic reasons explain why these constraints appear to work, mutual informations are excellent nonlinear correlation coefficients \cite{kinney2014equitability}, and it could be that high correlations are costly as some sort of intuitive memory cost.

\section{Derivation of a Generalized Blahut-Arimoto Algorithm}
\label{app:2}

We start with the unconstrained objective function
\begin{equation}
\mathcal{L} = \langle r(w_t,a_t)\rangle - \beta I[s_t;a_t] - \lambda I[\sigma_t;s_t] - \gamma_s \sum p(\sigma_t) p(s_t|\sigma_t) - \gamma_a \sum p(s_t) p(a_t|s_t)
\end{equation}
for discrete state spaces.
We take partial derivatives with respect to $p(a_t|s_t)$ and set them equal to $0$. First:
\begin{eqnarray}
\frac{\partial \langle r(w_t,a_t)\rangle}{\partial p(a_t|s_t)} &=& \frac{\partial}{\partial p(a_t|s_t)} \sum p(w_t,a_t) r(w_t,a_t) \\
&=& \frac{\partial}{\partial p(a_t|s_t)} \sum p(w_t,a_t,s_t,\sigma_t) r(w_t,a_t) \\
&=& \frac{\partial}{\partial p(a_t|s_t)} \sum p(a_t|s_t) p(s_t|\sigma_t) p(w_t|\sigma_t) p(\sigma_t) r(w_t,a_t) \\
&=& \sum p(\sigma_t) p(s_t|\sigma_t) p(w_t|\sigma_t) r(w_t,a_t).
\end{eqnarray}
Second:
\begin{eqnarray}
\frac{\partial I[s_t;a_t]}{\partial p(a_t|s_t)} &=& \frac{\partial}{\partial p(a_t|s_t)} \left(H[a_t]-H[a_t|s_t]\right)
\end{eqnarray}
where
\begin{eqnarray}
\frac{\partial H[a_t|s_t]}{\partial p(a_t|s_t)} &=& -\frac{\partial}{\partial p(a_t|s_t)} \sum p(s_t) p(a_t|s_t) \log p(a_t|s_t) \\
&=& -p(s_t) \left(1+\log p(a_t|s_t)\right)
\end{eqnarray}
and
\begin{eqnarray}
\frac{\partial H[a_t]}{\partial p(a_t|s_t)} &=& -\frac{\partial}{\partial p(a_t|s_t)} \sum p(a_t) \log p(a_t) \\
&=& -\sum (1+\log p(a)) \frac{\partial p(a)}{\partial p(a_t|s_t)} \\
&=& -\sum \delta_{a,a_t} p(s_t) (1+\log p(a)) \\
&=& -p(s_t) \left(1+ \log p(a_t)\right)
\end{eqnarray}
which means
\begin{eqnarray}
\frac{\partial I[s_t;a_t]}{\partial p(a_t|s_t)} &=& -p(s_t) \left(1+ \log p(a_t)\right)+p(s_t) \left(1+\log p(a_t|s_t)\right) \\
&=& p(s_t) \log \frac{p(a_t|s_t)}{p(a_t)}.
\end{eqnarray}
Third:
\begin{eqnarray}
\frac{\partial I[s_t;\sigma_t]}{\partial p(a_t|s_t)} &=& 0.
\end{eqnarray}
Fourth:
\begin{eqnarray}
\frac{\partial \sum p(a_t|s_t)}{\partial p(a_t|s_t)} &=& 1
\end{eqnarray}
and finally the last partial derivative is $0$. This gives
\begin{eqnarray}
0 &=& \sum_{\sigma_t,w_t} p(\sigma_t) p(s_t|\sigma_t) p(w_t|\sigma_t) r(w_t,a_t)-\beta p(s_t) \log \frac{p(a_t|s_t)}{p(a_t)} -\gamma_a p(s_t) \\
\beta p(s_t) \log \frac{p(a_t|s_t)}{p(a_t)} &=& \sum_{\sigma_t,w_t} p(\sigma_t) p(s_t|\sigma_t) p(w_t|\sigma_t) r(w_t,a_t) -\gamma_a p(s_t) \\
\log \frac{p(a_t|s_t)}{p(a_t)} &=& \frac{1}{\beta p(s_t)} \sum_{\sigma_t,w_t} p(\sigma_t) p(s_t|\sigma_t) p(w_t|\sigma_t) r(w_t,a_t) - \frac{\gamma_a}{\beta} \\
\frac{p(a_t|s_t)}{p(a_t)} &=& \exp\left(\frac{1}{p(s_t)} \sum_{\sigma_t,w_t} p(\sigma_t) p(s_t|\sigma_t) p(w_t|\sigma_t) r(w_t,a_t) - \frac{\gamma_a}{\beta}\right) \\
&=& \exp\left(\frac{1}{\beta}\sum_{\sigma_t,w_t} p(\sigma_t|s_t) p(w_t|\sigma_t) r(w_t,a_t) - \frac{\gamma_a}{\beta}\right) \\
p(a_t|s_t) &=& p(a_t) \frac{\exp \left(\frac{1}{\beta}\sum_{\sigma_t,w_t} p(\sigma_t|s_t) p(w_t|\sigma_t) r(w_t,a_t)\right)}{Z_{\beta}(s_t)}
\label{eq:1}
\end{eqnarray}
where $Z_{\beta}(a_t)$ is the partition function or normalization factor.
Similar manipulations for $p(s_t|\sigma_t)$ gives
\begin{equation}
p(s_t|\sigma_t) = \frac{p(s_t) \exp\left(\frac{1}{\lambda} \sum_{a_t,w_t} p(a_t|s_t) p(w_t|\sigma_t) r(w_t,a_t)\right)}{Z_{\lambda}(\sigma_t)}
\label{eq:2}
\end{equation}
where $Z_{\lambda}(\sigma_t)$ is the partition function or normalization factor. To retrieve the generalized Blahut-Arimoto algorithm for the two-dimensional rate-reward  manifold, we simply take Eqs. \ref{eq:1} and \ref{eq:2} and iterate them.

Every single time we iterate, we have to acknowledge that $p(w_t|\sigma_t)$ changes, as $p(a_t|s_t)$ and $p(s_t|\sigma_t)$ tell us how the action sequence changes. Hence, actually, $p(w|\sigma)$ is $p_n(w|\sigma)$, and $p(\sigma)$ is $p_n(\sigma)$, changing every iteration. How do we get these? A new action sequence, combined with the new observation sequence, tell us how the probability distribution over the world states changes. First note that
\begin{equation}
p(a|\sigma) = \sum_s p(a|s) p(s|\sigma)
\end{equation}
so that at each time step we have
\begin{eqnarray}
p(w_{t+1},o_t|w_t) &=& \sum_{a_t,\sigma_t} p(w_{t+1},o_t,a_t,\sigma_t|w_t) \\
&=& \sum_{a_t,\sigma_t} p(o_t|w_t) p(w_{t+1}|a_t,w_t) p(a_t|\sigma_t) p(\sigma_t)
\end{eqnarray}
which can be combined to make the labeled transition matrix $T^{(o_t)}$,
with which we can find the approximate probability distribution over the world states via the methods of Ref. \cite{crutchfield2016exact}:
\begin{equation}
p(w|\sigma) = p(w|\overleftarrow{o},\overleftarrow{a}) = \frac{\prod_t T^{(o_t)} \mu}{1^{\top}\prod_t T^{(o_t)}\mu}.
\end{equation}
Here, $\mu$ is the stationary distribution over world states, or the normalized $\text{eig}_1(\sum_o T^{(o)})$.
To get a rough estimate of this conditional probability distribution $p_{n+1}(w|\sigma)$ from this, we use a reasonably long observation sequence $\overleftarrow{o}^N$, making sure that $N$ is large enough to capture interesting behavior, though this encounters a curse of dimensionality if $N$ is too large \cite{marzen2016predictive}. It may seem as though the benefits of coarse-graining to sensorimotor causal states are lost by this maneuver, but now we only encounter a curse of dimensionality in finding $p(w|\sigma)$ and $p(\sigma)$ and not in finding $p(s|h)$. Each observation sequence leads to a different $\sigma$. We then use the methods of Ref. \cite{marzen2017nearly} to coarse-grain into approximate sensorimotor causal states to find $p_{n+1}(\sigma)$.

\section{Derivation of a minimax action policy when the lower-level organism stores sensorimotor causal states}
\label{app:3}

We start with
\begin{equation}
\pi_{minimax}(a|\sigma) = \arg\max_{\pi(a|\sigma)}\min_{p(w|\sigma)} \langle r(w,a)\rangle.
\end{equation}
and
\begin{equation}
\langle r(w,a)\rangle = \sum_{a,w,\sigma} p(\sigma) \pi(a|\sigma) p(w|\sigma) r(w,a).
\end{equation}
First we assume that $\pi(a|\sigma)$ is fixed and find the worst possible $p(w|\sigma)$:
\begin{eqnarray}
p_{minimax}(w|\sigma) &=& \min_{p(w|\sigma)} \sum_{a,w,\sigma} p(\sigma) \pi(a|\sigma) p(w|\sigma) r(w,a) - \sum_{w,\sigma} \lambda_{\sigma} p(w|\sigma)
\end{eqnarray}
so that
\begin{eqnarray}
0 &=& \sum_{a} p(\sigma) \pi(a|\sigma) r(w,a) - \lambda_{\sigma}.
\end{eqnarray}
The linearity in this objective implies that the objective is maximized at the edges of the simplex. In other words, $p(w|\sigma)$ should be $\delta_{w,f(\sigma)}$ for some $f(\sigma)$. In other words, there is a one-to-one mapping between sensorimotor causal states and hidden states, so that we might as well replace $\sigma$ with $w$ and assume that the environment is understood in this limit. Similar logic holds for $\pi(a|\sigma)$, so that $\pi(a|\sigma)$ is deterministic and $\delta_{a,g(w)}$. Altogether, this gives
\begin{equation}
    \pi_{minimax}(a|\sigma) = \delta_{a,g(w)}
\end{equation}
in which
\begin{equation}
g(w) = \arg\max_g \sum_{w} p_g(w) r(w,g(w)).
\end{equation}
To find $p_g(w)$, we use
\begin{eqnarray}
p_g(w_{t+1}) &=& \sum_{a_t,w_t} p(w_{t+1}|a_t,w_t) p(a_t|w_t) p_g(w_t) \\
&=& \sum_{a_t,w_t} p(w_{t+1}|a_t,w_t) \delta_{a_t,g(w_t)} p_g(w_t) \\
&=& \sum_{w_t} p(w_{t+1}|a_t=g(w_t),w_t) p_g(w_t)
\end{eqnarray}
so that
\begin{equation}
p_g(w) = \text{eig}_1(p(w'|a=g(w),w)). \label{eq:C10}
\end{equation}
We can do a brute force search and find the appropriate $g$. This gives us the following reward and rates:
\begin{eqnarray}
\langle r\rangle_{minimax} &=& \max_g \sum_{w} \text{eig}_1(p(w'|a=g(w),w)) r(w,g(w)) \\
I[s;h] &=& H[w] \\
I[a;s] &=& I[w,g(w)] = H[g(w)]. \label{eq:C13}
\end{eqnarray}
This point can then be placed next to the reward-rate manifold.

\section{Derivation of the example POMDP objective function}
\label{app:4}

We start by finding:
\begin{eqnarray}
s &=& m_{s,o} w + 0.01 m_{s,o}\eta_o + \sigma_{s,o}\eta_s \\
a &=& m_{a,s} s+\sigma_{a,s}\eta_a \\
&=& m_{a,s}m_{s,o}w+0.01 m_{a,s}m_{s,o}\eta_o+m_{a,s}\sigma_{s,o}\eta_s+\sigma_{a,s}\eta_a.
\end{eqnarray}
This implies that
\begin{equation}
w_{t+1} = w_t - m_{a,s}m_{s,o}w-0.01 m_{a,s}m_{s,o}\eta_o-m_{a,s}\sigma_{s,o}\eta_s-\sigma_{a,s}\eta_a+\eta_{w}
\end{equation}
where the noises all combine to make
\begin{equation}
w_{t+1} = (1-m_{a,s}m_{s,o})w_t + \sqrt{0.01^2 m_{a,s}^2m_{s,o}^2+m_{a,s}^2\sigma_{s,o}^2+\sigma_{a,s}^2+1} \eta_{comb}.
\end{equation}
This implies, if we find the variance of both sides, and assuming stationary statistics, that
\begin{equation}
\sigma_{ww} = (1-m_{a,s}m_{s,o})^2 \sigma_{ww} + \left(0.01^2 m_{a,s}^2m_{s,o}^2+m_{a,s}^2\sigma_{s,o}^2+\sigma_{a,s}^2+1\right)
\end{equation}
which means that
\begin{equation}
\sigma_{ww} = \frac{0.01^2 m_{a,s}^2m_{s,o}^2+m_{a,s}^2\sigma_{s,o}^2+\sigma_{a,s}^2+1}{1-(1-m_{a,s}m_{s,o})^2}.
\end{equation}

Several other quantities are necessary, as we need covariances and variances of other variables. We have
\begin{eqnarray}
\sigma_{ws} &=& \langle w_t s_t\rangle-\langle w_t\rangle\langle s_t\rangle \\
&=& m_{s,o}\sigma_{ww},
\end{eqnarray}
and similar calculations yield
\begin{eqnarray}
\sigma_{ss} &=& m_{s,o}^2 \sigma_{ww} + 0.01^2 m_{s,o}^2+\sigma_{s,o}^2 \\
\sigma_{as} &=& m_{a,s}m_{s,o}^2\sigma_{ww}+0.01^2 m_{a,s}m_{s,o}^2+m_{a,s}\sigma_{s,o}^2 \\
\sigma_{aa} &=& m_{a,s}^2 m_{s,o}^2 \sigma_{ww} + 0.01^2 m_{a,s}^2m_{s,o}^2+m_{a,s}^2\sigma_{s,o}^2+\sigma_{a,s}^2.
\end{eqnarray}
With the formula for the mutual information between two Gaussians, we have
\begin{eqnarray}
I[s;a] &=& - \frac{1}{2}\log \left(1-\frac{\sigma_{sa}^2}{\sigma_{ss}\sigma_{aa}}\right) \\
I[s;w] &=& -\frac{1}{2}\log \left(1-\frac{\sigma_{sw}^2}{\sigma_{ss}\sigma_{ww}}\right).
\end{eqnarray}
And then,
\begin{eqnarray}
\langle (w-a)^2\rangle &=& \langle \left((m_{a,s}m_{s,o} -1)w+\sqrt{0.01^2 m_{a,s}^2m_{s,o}^2+m_{a,s}^2\sigma_{s,o}^2+\sigma_{a,s}^2+1}\eta_{comb}\right)^2\rangle \\
&=& (m_{a,s}m_{s,o} -1)^2\sigma_{ww}+0.01^2 m_{a,s}^2m_{s,o}^2+m_{a,s}^2\sigma_{s,o}^2+\sigma_{a,s}^2+1.
\end{eqnarray}
The entire expression was loaded into Mathematica and numerically maximized for $\beta,\lambda$ ranging from $0$ to $1000$ with constraints that all variables were between $0$ and $1$ to avoid a nonstationarity that led to an unphysically negative variance for $w_t$.

\end{widetext}

\bibliography{apssamp}

\begin{thebibliography}{67}%
\makeatletter
\providecommand \@ifxundefined [1]{%
 \@ifx{#1\undefined}
}%
\providecommand \@ifnum [1]{%
 \ifnum #1\expandafter \@firstoftwo
 \else \expandafter \@secondoftwo
 \fi
}%
\providecommand \@ifx [1]{%
 \ifx #1\expandafter \@firstoftwo
 \else \expandafter \@secondoftwo
 \fi
}%
\providecommand \natexlab [1]{#1}%
\providecommand \enquote  [1]{``#1''}%
\providecommand \bibnamefont  [1]{#1}%
\providecommand \bibfnamefont [1]{#1}%
\providecommand \citenamefont [1]{#1}%
\providecommand \href@noop [0]{\@secondoftwo}%
\providecommand \href [0]{\begingroup \@sanitize@url \@href}%
\providecommand \@href[1]{\@@startlink{#1}\@@href}%
\providecommand \@@href[1]{\endgroup#1\@@endlink}%
\providecommand \@sanitize@url [0]{\catcode `\\12\catcode `\$12\catcode
  `\&12\catcode `\#12\catcode `\^12\catcode `\_12\catcode `\%12\relax}%
\providecommand \@@startlink[1]{}%
\providecommand \@@endlink[0]{}%
\providecommand \url  [0]{\begingroup\@sanitize@url \@url }%
\providecommand \@url [1]{\endgroup\@href {#1}{\urlprefix }}%
\providecommand \urlprefix  [0]{URL }%
\providecommand \Eprint [0]{\href }%
\providecommand \doibase [0]{https://doi.org/}%
\providecommand \selectlanguage [0]{\@gobble}%
\providecommand \bibinfo  [0]{\@secondoftwo}%
\providecommand \bibfield  [0]{\@secondoftwo}%
\providecommand \translation [1]{[#1]}%
\providecommand \BibitemOpen [0]{}%
\providecommand \bibitemStop [0]{}%
\providecommand \bibitemNoStop [0]{.\EOS\space}%
\providecommand \EOS [0]{\spacefactor3000\relax}%
\providecommand \BibitemShut  [1]{\csname bibitem#1\endcsname}%
\let\auto@bib@innerbib\@empty
\bibitem [{\citenamefont {Marr}(1982)}]{marr1982vision}%
  \BibitemOpen
  \bibfield  {author} {\bibinfo {author} {\bibfnamefont {D.}~\bibnamefont
  {Marr}},\ }\bibfield  {title} {\bibinfo {title} {Vision new york: Freeman},\
  }\href@noop {} {\  (\bibinfo {year} {1982})}\BibitemShut {NoStop}%
\bibitem [{\citenamefont {Levenstein}\ \emph {et~al.}(2023)\citenamefont
  {Levenstein}, \citenamefont {Alvarez}, \citenamefont {Amarasingham},
  \citenamefont {Azab}, \citenamefont {Chen}, \citenamefont {Gerkin},
  \citenamefont {Hasenstaub}, \citenamefont {Iyer}, \citenamefont {Jolivet},
  \citenamefont {Marzen} \emph {et~al.}}]{levenstein2023role}%
  \BibitemOpen
  \bibfield  {author} {\bibinfo {author} {\bibfnamefont {D.}~\bibnamefont
  {Levenstein}}, \bibinfo {author} {\bibfnamefont {V.~A.}\ \bibnamefont
  {Alvarez}}, \bibinfo {author} {\bibfnamefont {A.}~\bibnamefont
  {Amarasingham}}, \bibinfo {author} {\bibfnamefont {H.}~\bibnamefont {Azab}},
  \bibinfo {author} {\bibfnamefont {Z.~S.}\ \bibnamefont {Chen}}, \bibinfo
  {author} {\bibfnamefont {R.~C.}\ \bibnamefont {Gerkin}}, \bibinfo {author}
  {\bibfnamefont {A.}~\bibnamefont {Hasenstaub}}, \bibinfo {author}
  {\bibfnamefont {R.}~\bibnamefont {Iyer}}, \bibinfo {author} {\bibfnamefont
  {R.~B.}\ \bibnamefont {Jolivet}}, \bibinfo {author} {\bibfnamefont
  {S.}~\bibnamefont {Marzen}}, \emph {et~al.},\ }\bibfield  {title} {\bibinfo
  {title} {On the role of theory and modeling in neuroscience},\ }\href@noop {}
  {\bibfield  {journal} {\bibinfo  {journal} {Journal of Neuroscience}\
  }\textbf {\bibinfo {volume} {43}},\ \bibinfo {pages} {1074} (\bibinfo {year}
  {2023})}\BibitemShut {NoStop}%
\bibitem [{\citenamefont {Barlow}\ \emph {et~al.}(1961)\citenamefont {Barlow}
  \emph {et~al.}}]{barlow1961possible}%
  \BibitemOpen
  \bibfield  {author} {\bibinfo {author} {\bibfnamefont {H.~B.}\ \bibnamefont
  {Barlow}} \emph {et~al.},\ }\bibfield  {title} {\bibinfo {title} {Possible
  principles underlying the transformation of sensory messages},\ }\href@noop
  {} {\bibfield  {journal} {\bibinfo  {journal} {Sensory communication}\
  }\textbf {\bibinfo {volume} {1}},\ \bibinfo {pages} {217} (\bibinfo {year}
  {1961})}\BibitemShut {NoStop}%
\bibitem [{\citenamefont {Icard}(2023)}]{icard2023resource}%
  \BibitemOpen
  \bibfield  {author} {\bibinfo {author} {\bibfnamefont {T.~F.}\ \bibnamefont
  {Icard}},\ }\bibfield  {title} {\bibinfo {title} {Resource rationality},\
  }\href@noop {} {\  (\bibinfo {year} {2023})}\BibitemShut {NoStop}%
\bibitem [{\citenamefont {Gershman}\ \emph {et~al.}(2015)\citenamefont
  {Gershman}, \citenamefont {Horvitz},\ and\ \citenamefont
  {Tenenbaum}}]{gershman2015computational}%
  \BibitemOpen
  \bibfield  {author} {\bibinfo {author} {\bibfnamefont {S.~J.}\ \bibnamefont
  {Gershman}}, \bibinfo {author} {\bibfnamefont {E.~J.}\ \bibnamefont
  {Horvitz}},\ and\ \bibinfo {author} {\bibfnamefont {J.~B.}\ \bibnamefont
  {Tenenbaum}},\ }\bibfield  {title} {\bibinfo {title} {Computational
  rationality: A converging paradigm for intelligence in brains, minds, and
  machines},\ }\href@noop {} {\bibfield  {journal} {\bibinfo  {journal}
  {Science}\ }\textbf {\bibinfo {volume} {349}},\ \bibinfo {pages} {273}
  (\bibinfo {year} {2015})}\BibitemShut {NoStop}%
\bibitem [{\citenamefont {Sims}(2003)}]{sims2003implications}%
  \BibitemOpen
  \bibfield  {author} {\bibinfo {author} {\bibfnamefont {C.~A.}\ \bibnamefont
  {Sims}},\ }\bibfield  {title} {\bibinfo {title} {Implications of rational
  inattention},\ }\href@noop {} {\bibfield  {journal} {\bibinfo  {journal}
  {Journal of monetary Economics}\ }\textbf {\bibinfo {volume} {50}},\ \bibinfo
  {pages} {665} (\bibinfo {year} {2003})}\BibitemShut {NoStop}%
\bibitem [{\citenamefont {Sims}(2006)}]{sims2006rational}%
  \BibitemOpen
  \bibfield  {author} {\bibinfo {author} {\bibfnamefont {C.~A.}\ \bibnamefont
  {Sims}},\ }\bibfield  {title} {\bibinfo {title} {Rational inattention: Beyond
  the linear-quadratic case},\ }\href@noop {} {\bibfield  {journal} {\bibinfo
  {journal} {American Economic Review}\ }\textbf {\bibinfo {volume} {96}},\
  \bibinfo {pages} {158} (\bibinfo {year} {2006})}\BibitemShut {NoStop}%
\bibitem [{\citenamefont {Berger}(1971)}]{berger}%
  \BibitemOpen
  \bibfield  {author} {\bibinfo {author} {\bibfnamefont {T.}~\bibnamefont
  {Berger}},\ }\href@noop {} {\emph {\bibinfo {title} {Rate distortion theory:
  A mathematical basis for data compression}}}\ (\bibinfo  {publisher}
  {Prentice-Hall, Inc.},\ \bibinfo {year} {1971})\BibitemShut {NoStop}%
\bibitem [{\citenamefont {Sutton}\ and\ \citenamefont
  {Barto}(2018)}]{sutton2018reinforcement}%
  \BibitemOpen
  \bibfield  {author} {\bibinfo {author} {\bibfnamefont {R.~S.}\ \bibnamefont
  {Sutton}}\ and\ \bibinfo {author} {\bibfnamefont {A.~G.}\ \bibnamefont
  {Barto}},\ }\href@noop {} {\emph {\bibinfo {title} {Reinforcement learning:
  An introduction}}}\ (\bibinfo  {publisher} {MIT press},\ \bibinfo {year}
  {2018})\BibitemShut {NoStop}%
\bibitem [{\citenamefont {Schultz}\ \emph {et~al.}(1997)\citenamefont
  {Schultz}, \citenamefont {Dayan},\ and\ \citenamefont
  {Montague}}]{schultz1997neural}%
  \BibitemOpen
  \bibfield  {author} {\bibinfo {author} {\bibfnamefont {W.}~\bibnamefont
  {Schultz}}, \bibinfo {author} {\bibfnamefont {P.}~\bibnamefont {Dayan}},\
  and\ \bibinfo {author} {\bibfnamefont {P.~R.}\ \bibnamefont {Montague}},\
  }\bibfield  {title} {\bibinfo {title} {A neural substrate of prediction and
  reward},\ }\href@noop {} {\bibfield  {journal} {\bibinfo  {journal}
  {Science}\ }\textbf {\bibinfo {volume} {275}},\ \bibinfo {pages} {1593}
  (\bibinfo {year} {1997})}\BibitemShut {NoStop}%
\bibitem [{\citenamefont {Jeong}\ \emph {et~al.}(2022)\citenamefont {Jeong},
  \citenamefont {Taylor}, \citenamefont {Floeder}, \citenamefont {Lohmann},
  \citenamefont {Mihalas}, \citenamefont {Wu}, \citenamefont {Zhou},
  \citenamefont {Burke},\ and\ \citenamefont
  {Namboodiri}}]{jeong2022mesolimbic}%
  \BibitemOpen
  \bibfield  {author} {\bibinfo {author} {\bibfnamefont {H.}~\bibnamefont
  {Jeong}}, \bibinfo {author} {\bibfnamefont {A.}~\bibnamefont {Taylor}},
  \bibinfo {author} {\bibfnamefont {J.~R.}\ \bibnamefont {Floeder}}, \bibinfo
  {author} {\bibfnamefont {M.}~\bibnamefont {Lohmann}}, \bibinfo {author}
  {\bibfnamefont {S.}~\bibnamefont {Mihalas}}, \bibinfo {author} {\bibfnamefont
  {B.}~\bibnamefont {Wu}}, \bibinfo {author} {\bibfnamefont {M.}~\bibnamefont
  {Zhou}}, \bibinfo {author} {\bibfnamefont {D.~A.}\ \bibnamefont {Burke}},\
  and\ \bibinfo {author} {\bibfnamefont {V.~M.~K.}\ \bibnamefont
  {Namboodiri}},\ }\bibfield  {title} {\bibinfo {title} {Mesolimbic dopamine
  release conveys causal associations},\ }\href@noop {} {\bibfield  {journal}
  {\bibinfo  {journal} {Science}\ }\textbf {\bibinfo {volume} {378}},\ \bibinfo
  {pages} {eabq6740} (\bibinfo {year} {2022})}\BibitemShut {NoStop}%
\bibitem [{\citenamefont {Sims}(2018)}]{sims2018efficient}%
  \BibitemOpen
  \bibfield  {author} {\bibinfo {author} {\bibfnamefont {C.~R.}\ \bibnamefont
  {Sims}},\ }\bibfield  {title} {\bibinfo {title} {Efficient coding explains
  the universal law of generalization in human perception},\ }\href@noop {}
  {\bibfield  {journal} {\bibinfo  {journal} {Science}\ }\textbf {\bibinfo
  {volume} {360}},\ \bibinfo {pages} {652} (\bibinfo {year}
  {2018})}\BibitemShut {NoStop}%
\bibitem [{\citenamefont {Zaslavsky}\ \emph {et~al.}(2018)\citenamefont
  {Zaslavsky}, \citenamefont {Kemp}, \citenamefont {Regier},\ and\
  \citenamefont {Tishby}}]{zaslavsky2018efficient}%
  \BibitemOpen
  \bibfield  {author} {\bibinfo {author} {\bibfnamefont {N.}~\bibnamefont
  {Zaslavsky}}, \bibinfo {author} {\bibfnamefont {C.}~\bibnamefont {Kemp}},
  \bibinfo {author} {\bibfnamefont {T.}~\bibnamefont {Regier}},\ and\ \bibinfo
  {author} {\bibfnamefont {N.}~\bibnamefont {Tishby}},\ }\bibfield  {title}
  {\bibinfo {title} {Efficient compression in color naming and its evolution},\
  }\href@noop {} {\bibfield  {journal} {\bibinfo  {journal} {Proceedings of the
  National Academy of Sciences}\ }\textbf {\bibinfo {volume} {115}},\ \bibinfo
  {pages} {7937} (\bibinfo {year} {2018})}\BibitemShut {NoStop}%
\bibitem [{\citenamefont {Ferdinand}\ \emph {et~al.}(2024)\citenamefont
  {Ferdinand}, \citenamefont {Yu},\ and\ \citenamefont
  {Marzen}}]{humanbehavior}%
  \BibitemOpen
  \bibfield  {author} {\bibinfo {author} {\bibfnamefont {V.}~\bibnamefont
  {Ferdinand}}, \bibinfo {author} {\bibfnamefont {A.}~\bibnamefont {Yu}},\ and\
  \bibinfo {author} {\bibfnamefont {S.}~\bibnamefont {Marzen}},\ }\bibfield
  {title} {\bibinfo {title} {Humans are resource-rational predictors in a
  sequence learning task},\ }\href@noop {} {\bibfield  {journal} {\bibinfo
  {journal} {bioRxiv}\ ,\ \bibinfo {pages} {2024}} (\bibinfo {year}
  {2024})}\BibitemShut {NoStop}%
\bibitem [{\citenamefont {Jakob}\ and\ \citenamefont
  {Gershman}(2023)}]{jakob2023rate}%
  \BibitemOpen
  \bibfield  {author} {\bibinfo {author} {\bibfnamefont {A.~M.}\ \bibnamefont
  {Jakob}}\ and\ \bibinfo {author} {\bibfnamefont {S.~J.}\ \bibnamefont
  {Gershman}},\ }\bibfield  {title} {\bibinfo {title} {Rate-distortion theory
  of neural coding and its implications for working memory},\ }\href@noop {}
  {\bibfield  {journal} {\bibinfo  {journal} {Elife}\ }\textbf {\bibinfo
  {volume} {12}},\ \bibinfo {pages} {e79450} (\bibinfo {year}
  {2023})}\BibitemShut {NoStop}%
\bibitem [{\citenamefont {Palmer}\ \emph {et~al.}(2015)\citenamefont {Palmer},
  \citenamefont {Marre}, \citenamefont {Berry},\ and\ \citenamefont
  {Bialek}}]{palmer2015predictive}%
  \BibitemOpen
  \bibfield  {author} {\bibinfo {author} {\bibfnamefont {S.~E.}\ \bibnamefont
  {Palmer}}, \bibinfo {author} {\bibfnamefont {O.}~\bibnamefont {Marre}},
  \bibinfo {author} {\bibfnamefont {M.~J.}\ \bibnamefont {Berry}},\ and\
  \bibinfo {author} {\bibfnamefont {W.}~\bibnamefont {Bialek}},\ }\bibfield
  {title} {\bibinfo {title} {Predictive information in a sensory population},\
  }\href@noop {} {\bibfield  {journal} {\bibinfo  {journal} {Proceedings of the
  National Academy of Sciences}\ }\textbf {\bibinfo {volume} {112}},\ \bibinfo
  {pages} {6908} (\bibinfo {year} {2015})}\BibitemShut {NoStop}%
\bibitem [{\citenamefont {Lamberti}\ \emph
  {et~al.}(2023{\natexlab{a}})\citenamefont {Lamberti}, \citenamefont
  {Tripathi}, \citenamefont {van Putten}, \citenamefont {Marzen},\ and\
  \citenamefont {le~Feber}}]{pnasnexus}%
  \BibitemOpen
  \bibfield  {author} {\bibinfo {author} {\bibfnamefont {M.}~\bibnamefont
  {Lamberti}}, \bibinfo {author} {\bibfnamefont {S.}~\bibnamefont {Tripathi}},
  \bibinfo {author} {\bibfnamefont {M.~J. A.~M.}\ \bibnamefont {van Putten}},
  \bibinfo {author} {\bibfnamefont {S.}~\bibnamefont {Marzen}},\ and\ \bibinfo
  {author} {\bibfnamefont {J.}~\bibnamefont {le~Feber}},\ }\bibfield  {title}
  {\bibinfo {title} {Prediction in cultured cortical neural networks},\
  }\href@noop {} {\bibfield  {journal} {\bibinfo  {journal} {PNAS Nexus}\
  }\textbf {\bibinfo {volume} {2}} (\bibinfo {year}
  {2023}{\natexlab{a}})}\BibitemShut {NoStop}%
\bibitem [{\citenamefont {Hasenstaub}\ \emph {et~al.}(2010)\citenamefont
  {Hasenstaub}, \citenamefont {Otte}, \citenamefont {Callaway},\ and\
  \citenamefont {Sejnowski}}]{hasenstaub2010metabolic}%
  \BibitemOpen
  \bibfield  {author} {\bibinfo {author} {\bibfnamefont {A.}~\bibnamefont
  {Hasenstaub}}, \bibinfo {author} {\bibfnamefont {S.}~\bibnamefont {Otte}},
  \bibinfo {author} {\bibfnamefont {E.}~\bibnamefont {Callaway}},\ and\
  \bibinfo {author} {\bibfnamefont {T.~J.}\ \bibnamefont {Sejnowski}},\
  }\bibfield  {title} {\bibinfo {title} {Metabolic cost as a unifying principle
  governing neuronal biophysics},\ }\href@noop {} {\bibfield  {journal}
  {\bibinfo  {journal} {Proceedings of the National Academy of Sciences}\
  }\textbf {\bibinfo {volume} {107}},\ \bibinfo {pages} {12329} (\bibinfo
  {year} {2010})}\BibitemShut {NoStop}%
\bibitem [{\citenamefont {Mehta}\ and\ \citenamefont
  {Schwab}(2012)}]{mehta2012energetic}%
  \BibitemOpen
  \bibfield  {author} {\bibinfo {author} {\bibfnamefont {P.}~\bibnamefont
  {Mehta}}\ and\ \bibinfo {author} {\bibfnamefont {D.~J.}\ \bibnamefont
  {Schwab}},\ }\bibfield  {title} {\bibinfo {title} {Energetic costs of
  cellular computation},\ }\href@noop {} {\bibfield  {journal} {\bibinfo
  {journal} {Proceedings of the National Academy of Sciences}\ }\textbf
  {\bibinfo {volume} {109}},\ \bibinfo {pages} {17978} (\bibinfo {year}
  {2012})}\BibitemShut {NoStop}%
\bibitem [{\citenamefont {Still}(2009)}]{still2009information}%
  \BibitemOpen
  \bibfield  {author} {\bibinfo {author} {\bibfnamefont {S.}~\bibnamefont
  {Still}},\ }\bibfield  {title} {\bibinfo {title} {Information-theoretic
  approach to interactive learning},\ }\href@noop {} {\bibfield  {journal}
  {\bibinfo  {journal} {Europhysics Letters}\ }\textbf {\bibinfo {volume}
  {85}},\ \bibinfo {pages} {28005} (\bibinfo {year} {2009})}\BibitemShut
  {NoStop}%
\bibitem [{\citenamefont {Lai}\ and\ \citenamefont
  {Gershman}(2023)}]{lai2023human}%
  \BibitemOpen
  \bibfield  {author} {\bibinfo {author} {\bibfnamefont {L.}~\bibnamefont
  {Lai}}\ and\ \bibinfo {author} {\bibfnamefont {S.~J.}\ \bibnamefont
  {Gershman}},\ }\bibfield  {title} {\bibinfo {title} {Human decision making
  balances reward maximization and policy compression},\ }\href@noop {} {\
  (\bibinfo {year} {2023})}\BibitemShut {NoStop}%
\bibitem [{\citenamefont {Malloy}\ \emph {et~al.}(2021)\citenamefont {Malloy},
  \citenamefont {Sims}, \citenamefont {Klinger}, \citenamefont {Liu},
  \citenamefont {Riemer},\ and\ \citenamefont {Tesauro}}]{malloy2021capacity}%
  \BibitemOpen
  \bibfield  {author} {\bibinfo {author} {\bibfnamefont {T.}~\bibnamefont
  {Malloy}}, \bibinfo {author} {\bibfnamefont {C.~R.}\ \bibnamefont {Sims}},
  \bibinfo {author} {\bibfnamefont {T.}~\bibnamefont {Klinger}}, \bibinfo
  {author} {\bibfnamefont {M.}~\bibnamefont {Liu}}, \bibinfo {author}
  {\bibfnamefont {M.}~\bibnamefont {Riemer}},\ and\ \bibinfo {author}
  {\bibfnamefont {G.}~\bibnamefont {Tesauro}},\ }\bibfield  {title} {\bibinfo
  {title} {Capacity-limited decentralized actor-critic for multi-agent games},\
  }in\ \href@noop {} {\emph {\bibinfo {booktitle} {2021 IEEE Conference on
  Games (CoG)}}}\ (\bibinfo {organization} {IEEE},\ \bibinfo {year} {2021})\
  pp.\ \bibinfo {pages} {1--8}\BibitemShut {NoStop}%
\bibitem [{\citenamefont {Chechik}\ \emph {et~al.}(2003)\citenamefont
  {Chechik}, \citenamefont {Globerson}, \citenamefont {Tishby},\ and\
  \citenamefont {Weiss}}]{chechik2003information}%
  \BibitemOpen
  \bibfield  {author} {\bibinfo {author} {\bibfnamefont {G.}~\bibnamefont
  {Chechik}}, \bibinfo {author} {\bibfnamefont {A.}~\bibnamefont {Globerson}},
  \bibinfo {author} {\bibfnamefont {N.}~\bibnamefont {Tishby}},\ and\ \bibinfo
  {author} {\bibfnamefont {Y.}~\bibnamefont {Weiss}},\ }\bibfield  {title}
  {\bibinfo {title} {Information bottleneck for gaussian variables},\
  }\href@noop {} {\bibfield  {journal} {\bibinfo  {journal} {Advances in Neural
  Information Processing Systems}\ }\textbf {\bibinfo {volume} {16}} (\bibinfo
  {year} {2003})}\BibitemShut {NoStop}%
\bibitem [{\citenamefont {Bialek}\ \emph {et~al.}(2001)\citenamefont {Bialek},
  \citenamefont {Nemenman},\ and\ \citenamefont
  {Tishby}}]{bialek2001predictability}%
  \BibitemOpen
  \bibfield  {author} {\bibinfo {author} {\bibfnamefont {W.}~\bibnamefont
  {Bialek}}, \bibinfo {author} {\bibfnamefont {I.}~\bibnamefont {Nemenman}},\
  and\ \bibinfo {author} {\bibfnamefont {N.}~\bibnamefont {Tishby}},\
  }\bibfield  {title} {\bibinfo {title} {Predictability, complexity, and
  learning},\ }\href@noop {} {\bibfield  {journal} {\bibinfo  {journal} {Neural
  computation}\ }\textbf {\bibinfo {volume} {13}},\ \bibinfo {pages} {2409}
  (\bibinfo {year} {2001})}\BibitemShut {NoStop}%
\bibitem [{\citenamefont {Crutchfield}\ and\ \citenamefont
  {Marzen}(2015)}]{crutchfield2015signatures}%
  \BibitemOpen
  \bibfield  {author} {\bibinfo {author} {\bibfnamefont {J.~P.}\ \bibnamefont
  {Crutchfield}}\ and\ \bibinfo {author} {\bibfnamefont {S.}~\bibnamefont
  {Marzen}},\ }\bibfield  {title} {\bibinfo {title} {Signatures of infinity:
  Nonergodicity and resource scaling in prediction, complexity, and learning},\
  }\href@noop {} {\bibfield  {journal} {\bibinfo  {journal} {Physical Review
  E}\ }\textbf {\bibinfo {volume} {91}},\ \bibinfo {pages} {050106} (\bibinfo
  {year} {2015})}\BibitemShut {NoStop}%
\bibitem [{\citenamefont {Marzen}\ and\ \citenamefont
  {Crutchfield}(2016{\natexlab{a}})}]{marzen2016statistical}%
  \BibitemOpen
  \bibfield  {author} {\bibinfo {author} {\bibfnamefont {S.~E.}\ \bibnamefont
  {Marzen}}\ and\ \bibinfo {author} {\bibfnamefont {J.~P.}\ \bibnamefont
  {Crutchfield}},\ }\bibfield  {title} {\bibinfo {title} {Statistical
  signatures of structural organization: The case of long memory in renewal
  processes},\ }\href@noop {} {\bibfield  {journal} {\bibinfo  {journal}
  {Physics Letters A}\ }\textbf {\bibinfo {volume} {380}},\ \bibinfo {pages}
  {1517} (\bibinfo {year} {2016}{\natexlab{a}})}\BibitemShut {NoStop}%
\bibitem [{\citenamefont {Arumugam}\ \emph {et~al.}(2024)\citenamefont
  {Arumugam}, \citenamefont {Ho}, \citenamefont {Goodman},\ and\ \citenamefont
  {Van~Roy}}]{arumugam2024bayesian}%
  \BibitemOpen
  \bibfield  {author} {\bibinfo {author} {\bibfnamefont {D.}~\bibnamefont
  {Arumugam}}, \bibinfo {author} {\bibfnamefont {M.~K.}\ \bibnamefont {Ho}},
  \bibinfo {author} {\bibfnamefont {N.~D.}\ \bibnamefont {Goodman}},\ and\
  \bibinfo {author} {\bibfnamefont {B.}~\bibnamefont {Van~Roy}},\ }\bibfield
  {title} {\bibinfo {title} {Bayesian reinforcement learning with limited
  cognitive load},\ }\href@noop {} {\bibfield  {journal} {\bibinfo  {journal}
  {Open Mind}\ }\textbf {\bibinfo {volume} {8}},\ \bibinfo {pages} {395}
  (\bibinfo {year} {2024})}\BibitemShut {NoStop}%
\bibitem [{\citenamefont {Tucker}\ \emph {et~al.}(2022)\citenamefont {Tucker},
  \citenamefont {Levy}, \citenamefont {Shah},\ and\ \citenamefont
  {Zaslavsky}}]{NEURIPS2022_8bb5f663}%
  \BibitemOpen
  \bibfield  {author} {\bibinfo {author} {\bibfnamefont {M.}~\bibnamefont
  {Tucker}}, \bibinfo {author} {\bibfnamefont {R.}~\bibnamefont {Levy}},
  \bibinfo {author} {\bibfnamefont {J.~A.}\ \bibnamefont {Shah}},\ and\
  \bibinfo {author} {\bibfnamefont {N.}~\bibnamefont {Zaslavsky}},\ }\bibfield
  {title} {\bibinfo {title} {Trading off utility, informativeness, and
  complexity in emergent communication},\ }in\ \href
  {https://proceedings.neurips.cc/paper_files/paper/2022/file/8bb5f66371c7e4cbf6c223162c62c0f4-Paper-Conference.pdf}
  {\emph {\bibinfo {booktitle} {Advances in Neural Information Processing
  Systems}}},\ Vol.~\bibinfo {volume} {35},\ \bibinfo {editor} {edited by\
  \bibinfo {editor} {\bibfnamefont {S.}~\bibnamefont {Koyejo}}, \bibinfo
  {editor} {\bibfnamefont {S.}~\bibnamefont {Mohamed}}, \bibinfo {editor}
  {\bibfnamefont {A.}~\bibnamefont {Agarwal}}, \bibinfo {editor} {\bibfnamefont
  {D.}~\bibnamefont {Belgrave}}, \bibinfo {editor} {\bibfnamefont
  {K.}~\bibnamefont {Cho}},\ and\ \bibinfo {editor} {\bibfnamefont
  {A.}~\bibnamefont {Oh}}}\ (\bibinfo  {publisher} {Curran Associates, Inc.},\
  \bibinfo {year} {2022})\ pp.\ \bibinfo {pages} {22214--22228}\BibitemShut
  {NoStop}%
\bibitem [{\citenamefont {Tishby}\ and\ \citenamefont
  {Polani}(2010)}]{tishby2010information}%
  \BibitemOpen
  \bibfield  {author} {\bibinfo {author} {\bibfnamefont {N.}~\bibnamefont
  {Tishby}}\ and\ \bibinfo {author} {\bibfnamefont {D.}~\bibnamefont
  {Polani}},\ }\bibfield  {title} {\bibinfo {title} {Information theory of
  decisions and actions},\ }in\ \href@noop {} {\emph {\bibinfo {booktitle}
  {Perception-action cycle: Models, architectures, and hardware}}}\ (\bibinfo
  {publisher} {Springer},\ \bibinfo {year} {2010})\ pp.\ \bibinfo {pages}
  {601--636}\BibitemShut {NoStop}%
\bibitem [{\citenamefont {Van~Dijk}\ and\ \citenamefont
  {Polani}(2012)}]{van2012informational}%
  \BibitemOpen
  \bibfield  {author} {\bibinfo {author} {\bibfnamefont {S.~G.}\ \bibnamefont
  {Van~Dijk}}\ and\ \bibinfo {author} {\bibfnamefont {D.}~\bibnamefont
  {Polani}},\ }\bibfield  {title} {\bibinfo {title} {Informational drives for
  sensor evolution},\ }\href@noop {} {\bibfield  {journal} {\bibinfo  {journal}
  {Artificial Life 13}\ } (\bibinfo {year} {2012})}\BibitemShut {NoStop}%
\bibitem [{\citenamefont {Still}(2014)}]{still2014information}%
  \BibitemOpen
  \bibfield  {author} {\bibinfo {author} {\bibfnamefont {S.}~\bibnamefont
  {Still}},\ }\bibfield  {title} {\bibinfo {title} {Information bottleneck
  approach to predictive inference},\ }\href@noop {} {\bibfield  {journal}
  {\bibinfo  {journal} {Entropy}\ }\textbf {\bibinfo {volume} {16}},\ \bibinfo
  {pages} {968} (\bibinfo {year} {2014})}\BibitemShut {NoStop}%
\bibitem [{\citenamefont {Marzen}\ and\ \citenamefont
  {Crutchfield}(2016{\natexlab{b}})}]{marzen2016predictive}%
  \BibitemOpen
  \bibfield  {author} {\bibinfo {author} {\bibfnamefont {S.~E.}\ \bibnamefont
  {Marzen}}\ and\ \bibinfo {author} {\bibfnamefont {J.~P.}\ \bibnamefont
  {Crutchfield}},\ }\bibfield  {title} {\bibinfo {title} {Predictive
  rate-distortion for infinite-order markov processes},\ }\href@noop {}
  {\bibfield  {journal} {\bibinfo  {journal} {Journal of Statistical Physics}\
  }\textbf {\bibinfo {volume} {163}},\ \bibinfo {pages} {1312} (\bibinfo {year}
  {2016}{\natexlab{b}})}\BibitemShut {NoStop}%
\bibitem [{\citenamefont {Burnetas}\ and\ \citenamefont
  {Katehakis}(1996)}]{burnetas1996optimal}%
  \BibitemOpen
  \bibfield  {author} {\bibinfo {author} {\bibfnamefont {A.~N.}\ \bibnamefont
  {Burnetas}}\ and\ \bibinfo {author} {\bibfnamefont {M.~N.}\ \bibnamefont
  {Katehakis}},\ }\bibfield  {title} {\bibinfo {title} {Optimal adaptive
  policies for sequential allocation problems},\ }\href@noop {} {\bibfield
  {journal} {\bibinfo  {journal} {Advances in Applied Mathematics}\ }\textbf
  {\bibinfo {volume} {17}},\ \bibinfo {pages} {122} (\bibinfo {year}
  {1996})}\BibitemShut {NoStop}%
\bibitem [{\citenamefont {Chklovskii}\ and\ \citenamefont
  {Koulakov}(2004)}]{chklovskii2004maps}%
  \BibitemOpen
  \bibfield  {author} {\bibinfo {author} {\bibfnamefont {D.~B.}\ \bibnamefont
  {Chklovskii}}\ and\ \bibinfo {author} {\bibfnamefont {A.~A.}\ \bibnamefont
  {Koulakov}},\ }\bibfield  {title} {\bibinfo {title} {Maps in the brain: what
  can we learn from them?},\ }\href@noop {} {\bibfield  {journal} {\bibinfo
  {journal} {Annu. Rev. Neurosci.}\ }\textbf {\bibinfo {volume} {27}},\
  \bibinfo {pages} {369} (\bibinfo {year} {2004})}\BibitemShut {NoStop}%
\bibitem [{\citenamefont {Kaelbling}\ \emph {et~al.}(1998)\citenamefont
  {Kaelbling}, \citenamefont {Littman},\ and\ \citenamefont
  {Cassandra}}]{kaelbling1998planning}%
  \BibitemOpen
  \bibfield  {author} {\bibinfo {author} {\bibfnamefont {L.~P.}\ \bibnamefont
  {Kaelbling}}, \bibinfo {author} {\bibfnamefont {M.~L.}\ \bibnamefont
  {Littman}},\ and\ \bibinfo {author} {\bibfnamefont {A.~R.}\ \bibnamefont
  {Cassandra}},\ }\bibfield  {title} {\bibinfo {title} {Planning and acting in
  partially observable stochastic domains},\ }\href@noop {} {\bibfield
  {journal} {\bibinfo  {journal} {Artificial intelligence}\ }\textbf {\bibinfo
  {volume} {101}},\ \bibinfo {pages} {99} (\bibinfo {year} {1998})}\BibitemShut
  {NoStop}%
\bibitem [{\citenamefont {Doshi-Velez}\ \emph {et~al.}(2013)\citenamefont
  {Doshi-Velez}, \citenamefont {Pfau}, \citenamefont {Wood},\ and\
  \citenamefont {Roy}}]{doshi2013bayesian}%
  \BibitemOpen
  \bibfield  {author} {\bibinfo {author} {\bibfnamefont {F.}~\bibnamefont
  {Doshi-Velez}}, \bibinfo {author} {\bibfnamefont {D.}~\bibnamefont {Pfau}},
  \bibinfo {author} {\bibfnamefont {F.}~\bibnamefont {Wood}},\ and\ \bibinfo
  {author} {\bibfnamefont {N.}~\bibnamefont {Roy}},\ }\bibfield  {title}
  {\bibinfo {title} {Bayesian nonparametric methods for partially-observable
  reinforcement learning},\ }\href@noop {} {\bibfield  {journal} {\bibinfo
  {journal} {IEEE transactions on pattern analysis and machine intelligence}\
  }\textbf {\bibinfo {volume} {37}},\ \bibinfo {pages} {394} (\bibinfo {year}
  {2013})}\BibitemShut {NoStop}%
\bibitem [{\citenamefont {Shalizi}\ and\ \citenamefont
  {Crutchfield}(2001)}]{shalizi2001computational}%
  \BibitemOpen
  \bibfield  {author} {\bibinfo {author} {\bibfnamefont {C.~R.}\ \bibnamefont
  {Shalizi}}\ and\ \bibinfo {author} {\bibfnamefont {J.~P.}\ \bibnamefont
  {Crutchfield}},\ }\bibfield  {title} {\bibinfo {title} {Computational
  mechanics: Pattern and prediction, structure and simplicity},\ }\href@noop {}
  {\bibfield  {journal} {\bibinfo  {journal} {Journal of statistical physics}\
  }\textbf {\bibinfo {volume} {104}},\ \bibinfo {pages} {817} (\bibinfo {year}
  {2001})}\BibitemShut {NoStop}%
\bibitem [{\citenamefont {Tishby}\ \emph {et~al.}(2000)\citenamefont {Tishby},
  \citenamefont {Pereira},\ and\ \citenamefont
  {Bialek}}]{tishby2000information}%
  \BibitemOpen
  \bibfield  {author} {\bibinfo {author} {\bibfnamefont {N.}~\bibnamefont
  {Tishby}}, \bibinfo {author} {\bibfnamefont {F.~C.}\ \bibnamefont
  {Pereira}},\ and\ \bibinfo {author} {\bibfnamefont {W.}~\bibnamefont
  {Bialek}},\ }\bibfield  {title} {\bibinfo {title} {The information bottleneck
  method},\ }\href@noop {} {\bibfield  {journal} {\bibinfo  {journal} {arXiv
  preprint physics/0004057}\ } (\bibinfo {year} {2000})}\BibitemShut {NoStop}%
\bibitem [{\citenamefont {Moffett}\ and\ \citenamefont
  {Eckford}(2021)}]{moffett2021code}%
  \BibitemOpen
  \bibfield  {author} {\bibinfo {author} {\bibfnamefont {A.~S.}\ \bibnamefont
  {Moffett}}\ and\ \bibinfo {author} {\bibfnamefont {A.~W.}\ \bibnamefont
  {Eckford}},\ }\bibfield  {title} {\bibinfo {title} {To code, or not to code,
  at the racetrack: Kelly betting and single-letter codes},\ }\href@noop {}
  {\bibfield  {journal} {\bibinfo  {journal} {arXiv preprint arXiv:2104.14277}\
  } (\bibinfo {year} {2021})}\BibitemShut {NoStop}%
\bibitem [{\citenamefont {Marzen}(2024)}]{marzen2024comment}%
  \BibitemOpen
  \bibfield  {author} {\bibinfo {author} {\bibfnamefont {S.}~\bibnamefont
  {Marzen}},\ }\bibfield  {title} {\bibinfo {title} {Comment on deterministic
  information bottleneck},\ }\href@noop {} {\bibfield  {journal} {\bibinfo
  {journal} {arXiv preprint arXiv:2407.01786}\ } (\bibinfo {year}
  {2024})}\BibitemShut {NoStop}%
\bibitem [{\citenamefont {Crutchfield}\ \emph {et~al.}(2016)\citenamefont
  {Crutchfield}, \citenamefont {Ellison},\ and\ \citenamefont
  {Riechers}}]{crutchfield2016exact}%
  \BibitemOpen
  \bibfield  {author} {\bibinfo {author} {\bibfnamefont {J.~P.}\ \bibnamefont
  {Crutchfield}}, \bibinfo {author} {\bibfnamefont {C.~J.}\ \bibnamefont
  {Ellison}},\ and\ \bibinfo {author} {\bibfnamefont {P.~M.}\ \bibnamefont
  {Riechers}},\ }\bibfield  {title} {\bibinfo {title} {Exact complexity: The
  spectral decomposition of intrinsic computation},\ }\href@noop {} {\bibfield
  {journal} {\bibinfo  {journal} {Physics Letters A}\ }\textbf {\bibinfo
  {volume} {380}},\ \bibinfo {pages} {998} (\bibinfo {year}
  {2016})}\BibitemShut {NoStop}%
\bibitem [{\citenamefont {Marzen}\ and\ \citenamefont
  {Crutchfield}(2017)}]{marzen2017nearly}%
  \BibitemOpen
  \bibfield  {author} {\bibinfo {author} {\bibfnamefont {S.~E.}\ \bibnamefont
  {Marzen}}\ and\ \bibinfo {author} {\bibfnamefont {J.~P.}\ \bibnamefont
  {Crutchfield}},\ }\bibfield  {title} {\bibinfo {title} {Nearly maximally
  predictive features and their dimensions},\ }\href@noop {} {\bibfield
  {journal} {\bibinfo  {journal} {Physical Review E}\ }\textbf {\bibinfo
  {volume} {95}},\ \bibinfo {pages} {051301} (\bibinfo {year}
  {2017})}\BibitemShut {NoStop}%
\bibitem [{\citenamefont {Barnett}\ and\ \citenamefont
  {Crutchfield}(2015)}]{barnett2015computational}%
  \BibitemOpen
  \bibfield  {author} {\bibinfo {author} {\bibfnamefont {N.}~\bibnamefont
  {Barnett}}\ and\ \bibinfo {author} {\bibfnamefont {J.~P.}\ \bibnamefont
  {Crutchfield}},\ }\bibfield  {title} {\bibinfo {title} {Computational
  mechanics of input--output processes: Structured transformations and the
  epsilon-transducer},\ }\href@noop {} {\bibfield  {journal} {\bibinfo
  {journal} {Journal of Statistical Physics}\ }\textbf {\bibinfo {volume}
  {161}},\ \bibinfo {pages} {404} (\bibinfo {year} {2015})}\BibitemShut
  {NoStop}%
\bibitem [{\citenamefont {Celani}\ and\ \citenamefont
  {Vergassola}(2010)}]{celani2010bacterial}%
  \BibitemOpen
  \bibfield  {author} {\bibinfo {author} {\bibfnamefont {A.}~\bibnamefont
  {Celani}}\ and\ \bibinfo {author} {\bibfnamefont {M.}~\bibnamefont
  {Vergassola}},\ }\bibfield  {title} {\bibinfo {title} {Bacterial strategies
  for chemotaxis response},\ }\href@noop {} {\bibfield  {journal} {\bibinfo
  {journal} {Proceedings of the National Academy of Sciences}\ }\textbf
  {\bibinfo {volume} {107}},\ \bibinfo {pages} {1391} (\bibinfo {year}
  {2010})}\BibitemShut {NoStop}%
\bibitem [{\citenamefont {Mattingly}\ \emph {et~al.}(2021)\citenamefont
  {Mattingly}, \citenamefont {Kamino}, \citenamefont {Machta},\ and\
  \citenamefont {Emonet}}]{mattingly2021escherichia}%
  \BibitemOpen
  \bibfield  {author} {\bibinfo {author} {\bibfnamefont {H.}~\bibnamefont
  {Mattingly}}, \bibinfo {author} {\bibfnamefont {K.}~\bibnamefont {Kamino}},
  \bibinfo {author} {\bibfnamefont {B.}~\bibnamefont {Machta}},\ and\ \bibinfo
  {author} {\bibfnamefont {T.}~\bibnamefont {Emonet}},\ }\bibfield  {title}
  {\bibinfo {title} {Escherichia coli chemotaxis is information limited},\
  }\href@noop {} {\bibfield  {journal} {\bibinfo  {journal} {Nature physics}\
  }\textbf {\bibinfo {volume} {17}},\ \bibinfo {pages} {1426} (\bibinfo {year}
  {2021})}\BibitemShut {NoStop}%
\bibitem [{\citenamefont {Seifert}\ \emph {et~al.}(2024)\citenamefont
  {Seifert}, \citenamefont {Sealander}, \citenamefont {Marzen},\ and\
  \citenamefont {Levin}}]{seifert2024reinforcement}%
  \BibitemOpen
  \bibfield  {author} {\bibinfo {author} {\bibfnamefont {G.}~\bibnamefont
  {Seifert}}, \bibinfo {author} {\bibfnamefont {A.}~\bibnamefont {Sealander}},
  \bibinfo {author} {\bibfnamefont {S.}~\bibnamefont {Marzen}},\ and\ \bibinfo
  {author} {\bibfnamefont {M.}~\bibnamefont {Levin}},\ }\bibfield  {title}
  {\bibinfo {title} {From reinforcement learning to agency: Frameworks for
  understanding basal cognition},\ }\href@noop {} {\bibfield  {journal}
  {\bibinfo  {journal} {Biosystems}\ }\textbf {\bibinfo {volume} {235}},\
  \bibinfo {pages} {105107} (\bibinfo {year} {2024})}\BibitemShut {NoStop}%
\bibitem [{\citenamefont {Lamberti}\ \emph
  {et~al.}(2023{\natexlab{b}})\citenamefont {Lamberti}, \citenamefont
  {Tripathi}, \citenamefont {van Putten}, \citenamefont {Marzen},\ and\
  \citenamefont {le~Feber}}]{lamberti2023prediction}%
  \BibitemOpen
  \bibfield  {author} {\bibinfo {author} {\bibfnamefont {M.}~\bibnamefont
  {Lamberti}}, \bibinfo {author} {\bibfnamefont {S.}~\bibnamefont {Tripathi}},
  \bibinfo {author} {\bibfnamefont {M.~J.}\ \bibnamefont {van Putten}},
  \bibinfo {author} {\bibfnamefont {S.}~\bibnamefont {Marzen}},\ and\ \bibinfo
  {author} {\bibfnamefont {J.}~\bibnamefont {le~Feber}},\ }\bibfield  {title}
  {\bibinfo {title} {Prediction in cultured cortical neural networks},\
  }\href@noop {} {\bibfield  {journal} {\bibinfo  {journal} {PNAS nexus}\
  }\textbf {\bibinfo {volume} {2}},\ \bibinfo {pages} {pgad188} (\bibinfo
  {year} {2023}{\natexlab{b}})}\BibitemShut {NoStop}%
\bibitem [{\citenamefont {Hahn}\ and\ \citenamefont
  {Futrell}(2019)}]{hahn2019estimating}%
  \BibitemOpen
  \bibfield  {author} {\bibinfo {author} {\bibfnamefont {M.}~\bibnamefont
  {Hahn}}\ and\ \bibinfo {author} {\bibfnamefont {R.}~\bibnamefont {Futrell}},\
  }\bibfield  {title} {\bibinfo {title} {Estimating predictive rate--distortion
  curves via neural variational inference},\ }\href@noop {} {\bibfield
  {journal} {\bibinfo  {journal} {Entropy}\ }\textbf {\bibinfo {volume} {21}},\
  \bibinfo {pages} {640} (\bibinfo {year} {2019})}\BibitemShut {NoStop}%
\bibitem [{\citenamefont {Alemi}(2020)}]{alemi2020variational}%
  \BibitemOpen
  \bibfield  {author} {\bibinfo {author} {\bibfnamefont {A.~A.}\ \bibnamefont
  {Alemi}},\ }\bibfield  {title} {\bibinfo {title} {Variational predictive
  information bottleneck},\ }in\ \href@noop {} {\emph {\bibinfo {booktitle}
  {Symposium on Advances in Approximate Bayesian Inference}}}\ (\bibinfo
  {organization} {PMLR},\ \bibinfo {year} {2020})\ pp.\ \bibinfo {pages}
  {1--6}\BibitemShut {NoStop}%
\bibitem [{\citenamefont {Alemi}\ \emph {et~al.}(2016)\citenamefont {Alemi},
  \citenamefont {Fischer}, \citenamefont {Dillon},\ and\ \citenamefont
  {Murphy}}]{alemi2016deep}%
  \BibitemOpen
  \bibfield  {author} {\bibinfo {author} {\bibfnamefont {A.~A.}\ \bibnamefont
  {Alemi}}, \bibinfo {author} {\bibfnamefont {I.}~\bibnamefont {Fischer}},
  \bibinfo {author} {\bibfnamefont {J.~V.}\ \bibnamefont {Dillon}},\ and\
  \bibinfo {author} {\bibfnamefont {K.}~\bibnamefont {Murphy}},\ }\bibfield
  {title} {\bibinfo {title} {Deep variational information bottleneck},\
  }\href@noop {} {\bibfield  {journal} {\bibinfo  {journal} {arXiv preprint
  arXiv:1612.00410}\ } (\bibinfo {year} {2016})}\BibitemShut {NoStop}%
\bibitem [{\citenamefont {Kinney}\ and\ \citenamefont
  {Atwal}(2014)}]{kinney2014equitability}%
  \BibitemOpen
  \bibfield  {author} {\bibinfo {author} {\bibfnamefont {J.~B.}\ \bibnamefont
  {Kinney}}\ and\ \bibinfo {author} {\bibfnamefont {G.~S.}\ \bibnamefont
  {Atwal}},\ }\bibfield  {title} {\bibinfo {title} {Equitability, mutual
  information, and the maximal information coefficient},\ }\href@noop {}
  {\bibfield  {journal} {\bibinfo  {journal} {Proceedings of the National
  Academy of Sciences}\ }\textbf {\bibinfo {volume} {111}},\ \bibinfo {pages}
  {3354} (\bibinfo {year} {2014})}\BibitemShut {NoStop}%
\bibitem [{\citenamefont {Sridhar}\ \emph {et~al.}(2023)\citenamefont
  {Sridhar}, \citenamefont {Khamaj},\ and\ \citenamefont
  {Asthana}}]{sridhar2023cognitive}%
  \BibitemOpen
  \bibfield  {author} {\bibinfo {author} {\bibfnamefont {S.}~\bibnamefont
  {Sridhar}}, \bibinfo {author} {\bibfnamefont {A.}~\bibnamefont {Khamaj}},\
  and\ \bibinfo {author} {\bibfnamefont {M.~K.}\ \bibnamefont {Asthana}},\
  }\bibfield  {title} {\bibinfo {title} {Cognitive neuroscience perspective on
  memory: overview and summary},\ }\href@noop {} {\bibfield  {journal}
  {\bibinfo  {journal} {Frontiers in Human Neuroscience}\ }\textbf {\bibinfo
  {volume} {17}},\ \bibinfo {pages} {1217093} (\bibinfo {year}
  {2023})}\BibitemShut {NoStop}%
\bibitem [{\citenamefont {Still}\ \emph {et~al.}(2012)\citenamefont {Still},
  \citenamefont {Sivak}, \citenamefont {Bell},\ and\ \citenamefont
  {Crooks}}]{still2012thermodynamics}%
  \BibitemOpen
  \bibfield  {author} {\bibinfo {author} {\bibfnamefont {S.}~\bibnamefont
  {Still}}, \bibinfo {author} {\bibfnamefont {D.~A.}\ \bibnamefont {Sivak}},
  \bibinfo {author} {\bibfnamefont {A.~J.}\ \bibnamefont {Bell}},\ and\
  \bibinfo {author} {\bibfnamefont {G.~E.}\ \bibnamefont {Crooks}},\ }\bibfield
   {title} {\bibinfo {title} {Thermodynamics of prediction},\ }\href@noop {}
  {\bibfield  {journal} {\bibinfo  {journal} {Physical review letters}\
  }\textbf {\bibinfo {volume} {109}},\ \bibinfo {pages} {120604} (\bibinfo
  {year} {2012})}\BibitemShut {NoStop}%
\bibitem [{\citenamefont {Marzen}\ and\ \citenamefont
  {Crutchfield}(2020)}]{marzen2020prediction}%
  \BibitemOpen
  \bibfield  {author} {\bibinfo {author} {\bibfnamefont {S.~E.}\ \bibnamefont
  {Marzen}}\ and\ \bibinfo {author} {\bibfnamefont {J.~P.}\ \bibnamefont
  {Crutchfield}},\ }\bibfield  {title} {\bibinfo {title} {Prediction and
  dissipation in nonequilibrium molecular sensors: Conditionally markovian
  channels driven by memoryful environments},\ }\href@noop {} {\bibfield
  {journal} {\bibinfo  {journal} {Bulletin of mathematical biology}\ }\textbf
  {\bibinfo {volume} {82}},\ \bibinfo {pages} {25} (\bibinfo {year}
  {2020})}\BibitemShut {NoStop}%
\bibitem [{\citenamefont {Lan}\ \emph {et~al.}(2012)\citenamefont {Lan},
  \citenamefont {Sartori}, \citenamefont {Neumann}, \citenamefont {Sourjik},\
  and\ \citenamefont {Tu}}]{lan2012energy}%
  \BibitemOpen
  \bibfield  {author} {\bibinfo {author} {\bibfnamefont {G.}~\bibnamefont
  {Lan}}, \bibinfo {author} {\bibfnamefont {P.}~\bibnamefont {Sartori}},
  \bibinfo {author} {\bibfnamefont {S.}~\bibnamefont {Neumann}}, \bibinfo
  {author} {\bibfnamefont {V.}~\bibnamefont {Sourjik}},\ and\ \bibinfo {author}
  {\bibfnamefont {Y.}~\bibnamefont {Tu}},\ }\bibfield  {title} {\bibinfo
  {title} {The energy--speed--accuracy trade-off in sensory adaptation},\
  }\href@noop {} {\bibfield  {journal} {\bibinfo  {journal} {Nature physics}\
  }\textbf {\bibinfo {volume} {8}},\ \bibinfo {pages} {422} (\bibinfo {year}
  {2012})}\BibitemShut {NoStop}%
\bibitem [{\citenamefont {Moskovitz}\ \emph {et~al.}(2024)\citenamefont
  {Moskovitz}, \citenamefont {Miller}, \citenamefont {Sahani},\ and\
  \citenamefont {Botvinick}}]{moskovitz2024understanding}%
  \BibitemOpen
  \bibfield  {author} {\bibinfo {author} {\bibfnamefont {T.}~\bibnamefont
  {Moskovitz}}, \bibinfo {author} {\bibfnamefont {K.~J.}\ \bibnamefont
  {Miller}}, \bibinfo {author} {\bibfnamefont {M.}~\bibnamefont {Sahani}},\
  and\ \bibinfo {author} {\bibfnamefont {M.~M.}\ \bibnamefont {Botvinick}},\
  }\bibfield  {title} {\bibinfo {title} {Understanding dual process cognition
  via the minimum description length principle},\ }\href@noop {} {\bibfield
  {journal} {\bibinfo  {journal} {PLOS Computational Biology}\ }\textbf
  {\bibinfo {volume} {20}},\ \bibinfo {pages} {e1012383} (\bibinfo {year}
  {2024})}\BibitemShut {NoStop}%
\bibitem [{\citenamefont {Uppal}\ \emph {et~al.}(2020)\citenamefont {Uppal},
  \citenamefont {Ferdinand},\ and\ \citenamefont
  {Marzen}}]{uppal2020inferring}%
  \BibitemOpen
  \bibfield  {author} {\bibinfo {author} {\bibfnamefont {A.}~\bibnamefont
  {Uppal}}, \bibinfo {author} {\bibfnamefont {V.}~\bibnamefont {Ferdinand}},\
  and\ \bibinfo {author} {\bibfnamefont {S.}~\bibnamefont {Marzen}},\
  }\bibfield  {title} {\bibinfo {title} {Inferring an observer’s prediction
  strategy in sequence learning experiments},\ }\href@noop {} {\bibfield
  {journal} {\bibinfo  {journal} {Entropy}\ }\textbf {\bibinfo {volume} {22}},\
  \bibinfo {pages} {896} (\bibinfo {year} {2020})}\BibitemShut {NoStop}%
\bibitem [{\citenamefont {Daw}\ \emph {et~al.}(2011)\citenamefont {Daw} \emph
  {et~al.}}]{daw2011trial}%
  \BibitemOpen
  \bibfield  {author} {\bibinfo {author} {\bibfnamefont {N.~D.}\ \bibnamefont
  {Daw}} \emph {et~al.},\ }\bibfield  {title} {\bibinfo {title} {Trial-by-trial
  data analysis using computational models},\ }\href@noop {} {\bibfield
  {journal} {\bibinfo  {journal} {Decision making, affect, and learning:
  Attention and performance XXIII}\ }\textbf {\bibinfo {volume} {23}} (\bibinfo
  {year} {2011})}\BibitemShut {NoStop}%
\bibitem [{\citenamefont {Lai}\ and\ \citenamefont
  {Gershman}(2024)}]{lai2024human}%
  \BibitemOpen
  \bibfield  {author} {\bibinfo {author} {\bibfnamefont {L.}~\bibnamefont
  {Lai}}\ and\ \bibinfo {author} {\bibfnamefont {S.~J.}\ \bibnamefont
  {Gershman}},\ }\bibfield  {title} {\bibinfo {title} {Human decision making
  balances reward maximization and policy compression},\ }\href@noop {}
  {\bibfield  {journal} {\bibinfo  {journal} {PLOS Computational Biology}\
  }\textbf {\bibinfo {volume} {20}},\ \bibinfo {pages} {e1012057} (\bibinfo
  {year} {2024})}\BibitemShut {NoStop}%
\bibitem [{\citenamefont {Tka{\v{c}}ik}\ and\ \citenamefont
  {Wolde}(2025)}]{tkavcik2025information}%
  \BibitemOpen
  \bibfield  {author} {\bibinfo {author} {\bibfnamefont {G.}~\bibnamefont
  {Tka{\v{c}}ik}}\ and\ \bibinfo {author} {\bibfnamefont {P.~R.~t.}\
  \bibnamefont {Wolde}},\ }\bibfield  {title} {\bibinfo {title} {Information
  processing in biochemical networks},\ }\href@noop {} {\bibfield  {journal}
  {\bibinfo  {journal} {Annual Review of Biophysics}\ }\textbf {\bibinfo
  {volume} {54}} (\bibinfo {year} {2025})}\BibitemShut {NoStop}%
\bibitem [{\citenamefont {Laughlin}(1981)}]{laughlin1981simple}%
  \BibitemOpen
  \bibfield  {author} {\bibinfo {author} {\bibfnamefont {S.}~\bibnamefont
  {Laughlin}},\ }\bibfield  {title} {\bibinfo {title} {A simple coding
  procedure enhances a neuron's information capacity},\ }\href@noop {}
  {\bibfield  {journal} {\bibinfo  {journal} {Zeitschrift f{\"u}r
  Naturforschung c}\ }\textbf {\bibinfo {volume} {36}},\ \bibinfo {pages} {910}
  (\bibinfo {year} {1981})}\BibitemShut {NoStop}%
\bibitem [{\citenamefont {Park}\ and\ \citenamefont
  {Pillow}(2017)}]{park2017bayesian}%
  \BibitemOpen
  \bibfield  {author} {\bibinfo {author} {\bibfnamefont {I.~M.}\ \bibnamefont
  {Park}}\ and\ \bibinfo {author} {\bibfnamefont {J.~W.}\ \bibnamefont
  {Pillow}},\ }\bibfield  {title} {\bibinfo {title} {Bayesian efficient
  coding},\ }\href@noop {} {\bibfield  {journal} {\bibinfo  {journal}
  {BioRxiv}\ ,\ \bibinfo {pages} {178418}} (\bibinfo {year}
  {2017})}\BibitemShut {NoStop}%
\bibitem [{\citenamefont {Kostina}\ and\ \citenamefont
  {Verd{\'u}}(2012)}]{kostina2012fixed}%
  \BibitemOpen
  \bibfield  {author} {\bibinfo {author} {\bibfnamefont {V.}~\bibnamefont
  {Kostina}}\ and\ \bibinfo {author} {\bibfnamefont {S.}~\bibnamefont
  {Verd{\'u}}},\ }\bibfield  {title} {\bibinfo {title} {Fixed-length lossy
  compression in the finite blocklength regime},\ }\href@noop {} {\bibfield
  {journal} {\bibinfo  {journal} {IEEE Transactions on Information Theory}\
  }\textbf {\bibinfo {volume} {58}},\ \bibinfo {pages} {3309} (\bibinfo {year}
  {2012})}\BibitemShut {NoStop}%
\bibitem [{\citenamefont {Nemenman}\ \emph {et~al.}(2008)\citenamefont
  {Nemenman}, \citenamefont {Lewen}, \citenamefont {Bialek},\ and\
  \citenamefont {de~Ruyter~van Steveninck}}]{nemenman2008neural}%
  \BibitemOpen
  \bibfield  {author} {\bibinfo {author} {\bibfnamefont {I.}~\bibnamefont
  {Nemenman}}, \bibinfo {author} {\bibfnamefont {G.~D.}\ \bibnamefont {Lewen}},
  \bibinfo {author} {\bibfnamefont {W.}~\bibnamefont {Bialek}},\ and\ \bibinfo
  {author} {\bibfnamefont {R.~R.}\ \bibnamefont {de~Ruyter~van Steveninck}},\
  }\bibfield  {title} {\bibinfo {title} {Neural coding of natural stimuli:
  information at sub-millisecond resolution},\ }\href@noop {} {\bibfield
  {journal} {\bibinfo  {journal} {PLoS computational biology}\ }\textbf
  {\bibinfo {volume} {4}},\ \bibinfo {pages} {e1000025} (\bibinfo {year}
  {2008})}\BibitemShut {NoStop}%
\bibitem [{\citenamefont {Sawaya}\ \emph {et~al.}(2023)\citenamefont {Sawaya},
  \citenamefont {Issa},\ and\ \citenamefont {Marzen}}]{sawaya2023framework}%
  \BibitemOpen
  \bibfield  {author} {\bibinfo {author} {\bibfnamefont {Y.}~\bibnamefont
  {Sawaya}}, \bibinfo {author} {\bibfnamefont {G.}~\bibnamefont {Issa}},\ and\
  \bibinfo {author} {\bibfnamefont {S.~E.}\ \bibnamefont {Marzen}},\ }\bibfield
   {title} {\bibinfo {title} {Framework for solving time-delayed markov
  decision processes},\ }\href@noop {} {\bibfield  {journal} {\bibinfo
  {journal} {Physical Review Research}\ }\textbf {\bibinfo {volume} {5}},\
  \bibinfo {pages} {033034} (\bibinfo {year} {2023})}\BibitemShut {NoStop}%
\bibitem [{\citenamefont {Arnold}\ \emph {et~al.}(2006)\citenamefont {Arnold},
  \citenamefont {Loeliger}, \citenamefont {Vontobel}, \citenamefont {Kavcic},\
  and\ \citenamefont {Zeng}}]{arnold2006simulation}%
  \BibitemOpen
  \bibfield  {author} {\bibinfo {author} {\bibfnamefont {D.-M.}\ \bibnamefont
  {Arnold}}, \bibinfo {author} {\bibfnamefont {H.-A.}\ \bibnamefont
  {Loeliger}}, \bibinfo {author} {\bibfnamefont {P.~O.}\ \bibnamefont
  {Vontobel}}, \bibinfo {author} {\bibfnamefont {A.}~\bibnamefont {Kavcic}},\
  and\ \bibinfo {author} {\bibfnamefont {W.}~\bibnamefont {Zeng}},\ }\bibfield
  {title} {\bibinfo {title} {Simulation-based computation of information rates
  for channels with memory},\ }\href@noop {} {\bibfield  {journal} {\bibinfo
  {journal} {IEEE Transactions on information theory}\ }\textbf {\bibinfo
  {volume} {52}},\ \bibinfo {pages} {3498} (\bibinfo {year}
  {2006})}\BibitemShut {NoStop}%
\bibitem [{\citenamefont {Sagawa}\ and\ \citenamefont
  {Ueda}(2009)}]{sagawa2009minimal}%
  \BibitemOpen
  \bibfield  {author} {\bibinfo {author} {\bibfnamefont {T.}~\bibnamefont
  {Sagawa}}\ and\ \bibinfo {author} {\bibfnamefont {M.}~\bibnamefont {Ueda}},\
  }\bibfield  {title} {\bibinfo {title} {Minimal energy cost for thermodynamic
  information processing: measurement and information erasure},\ }\href@noop {}
  {\bibfield  {journal} {\bibinfo  {journal} {Physical review letters}\
  }\textbf {\bibinfo {volume} {102}},\ \bibinfo {pages} {250602} (\bibinfo
  {year} {2009})}\BibitemShut {NoStop}%
\end{thebibliography}%

\end{document}